# MAPPING THE HUMAN BRAIN FROM THE PRENATAL PERIOD TO INFANCY USING 3D MAGNETIC RESONANCE IMAGING: CORTICAL FOLDING AND EARLY GREY AND WHITE MATURATION PROCESSES

**Arnaud Cachia[1,2,3], Jean-François Mangin[4], and Jessica Dubois[5,6]**

1. Université de Paris, LaPsyDÉ, CNRS, F-75005 Paris, France
2. Université de Paris, IPNP, INSERM, F-75005 Paris, France.
3. Institut Universitaire de France, Paris, France
4. Université Paris-Saclay, CEA, CNRS, Neurospin, Baobab, Gif-sur-Yvette, F-91191, France
5. Université de Paris, NeuroDiderot, INSERM, F-75019 Paris, France
6. CEA, NeuroSpin, UNIACT, F-91191 Gif-sur-Yvette, France

# 1. Introduction

Human brain development is a complex and dynamic process that begins during the first weeks of pregnancy and lasts until early adulthood. This chapter will focus on the developmental window from prenatal period to infancy, probably the most dynamic period across the entire lifespan. The availability of non-invasive three-dimensional Magnetic Resonance Imaging (MRI) methodologies has changed the paradigm and allows investigations of the living human brain structure - e.g. micro- and macrostructural features of cortical and subcortical regions and their connections, including cortical sulcation/gyrification, area, and thickness, as well as white matter microstructure and connectivity, *see Box 1-3* - beginning *in utero*. Because of its relative safety, MRI is well-adapted to study individuals at multiple time points and to longitudinally follow the changes in brain structure and function that underlie the early stages of cognitive development.

Before the advent of brain imaging tools, structural brain changes were inferred from *post mortem* data with major concerns about their generalizability due to the questionable good health of the studied individuals with very young death. On the other hand, despite the advantages of *in vivo* MRI described earlier, MRI cannot measure structural changes at the cellular or molecular levels, and the physiological interpretation of MRI signal is not straightforward nor univocal. A detailed description of early developmental mechanisms (e.g. formation of the neural tube, neuronal migration and differentiation), as well as the early brain organization in transitory compartments such as the subplate can be found in complementary reviews (Kostovic & Judas, 2015; Stiles & Jernigan, 2010).

In addition, MRI data acquired on the early stage of brain development are very noisy due to the acquisition constraints (e.g. fetus movements, short acquisition duration) and difficult to analyze due to the low spatial resolution and the age-dependent tissue contrast and structure size. The comparison of anatomical brain measures derived from MRI across ages should therefore remain cautious. However, despite these limitations, the spatial and temporal patterns of developmental changes observed in recent MRI studies reflect patterns that were observed *post mortem*, demonstrating the validity and compatibility of these methods (Dehaene-Lambertz & Spelke, 2015; Dubois & Dehaene-Lambertz, 2015; Dubois, Kostovic, & Judas, 2015)

We here review how the cortex grows and gets convoluted, the microstructural maturation of the gray and white matter and their cognitive correlates in normal condition (see (Anderson, Spencer-Smith, & Wood, 2011; Miller, Huppi, & Mallard, 2016) in pathological conditions).

## 2. The early development of brain cortex

### *2.1. Cortical volume*

The last weeks of pregnancy and the first postnatal months are marked by an intense increase in cortical volume (Dubois & Dehaene-Lambertz, 2015) which progressively slows down after 2 years of age until adolescence. Changes in cortical volume are driven by changes in cortical thickness (CT) and cortical surface area (CSA), two complementary macrostructural features of the cortex anatomy with distinct genetic (Panizzon et al., 2009; Raznahan et al., 2011) and developmental (Raznahan et al., 2011) mechanisms. The classical “radial unit hypothesis” (Rakic, 1988, 2000) assumes that the CSA of a given cortical area essentially reflects the number of cortical columns, while CT essentially reflects the number and size of cells within a column and packing density, as well as the number of connections and the extent of their myelination (Eickhoff et al., 2005).

During fetal life, the total cortical volume increases from 10 $cm^3$ at 18 weeks of gestational age (wGA), to 30 $cm^3$ at 27 wGA, 150 $cm^3$ at 39 wGA (Andescavage et al., 2017; Makropoulos et al., 2016) reaching 200 $cm^3$ one month after birth and 600 $cm^3$ at two years old (Knickmeyer et al., 2008) (Figure 1a). The total CSA of fetus brain at 27 wGA is around 150 $cm^2$ (Makropoulos et al., 2016), reaching 600-800 $cm^2$ one month after birth, 1300 $cm^2$ at 5 months and 2000 $cm^2$ at 24 months (Dubois et al., 2019; Lyall et al., 2015).

The cortical volume increase is more important during the first postnatal year (~100%) than during the second year (~20%) (Gilmore et al., 2012). Developmental rates differ among brain regions, with important cortical volume increases in parietal and occipital regions in utero (Rajagopalan et al., 2011) and in association cortices, particularly in the frontal and parietal lobes, after birth (Gilmore et al., 2012). During the first two years after birth, brain growth is mainly due to gray matter development (Gilmore et al., 2007; Knickmeyer et al., 2008), which is no more the case later (Matsuzawa et al., 2001).

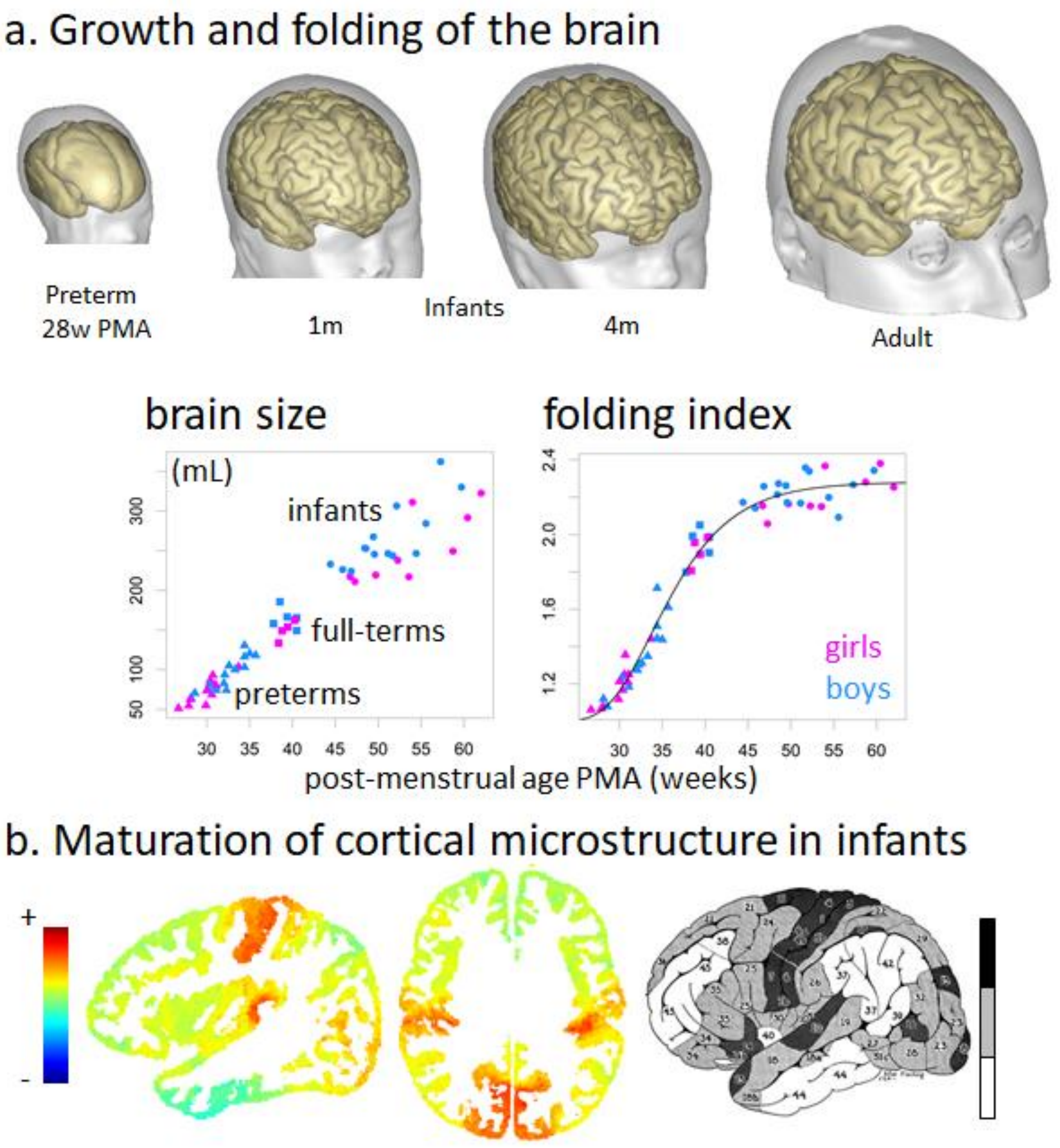

***Figure 1: Structural changes of the brain during development.*** *a: The brain shows intense growth and folding during the last weeks of pregnancy and the first months of infancy, as demonstrated with anatomical MRI. Illustrations are provided for a preterm newborn at 28 weeks of post-menstrual age (PMA), infants (at 1 and 4 months after birth) and an adult. Measures of brain size and folding index as a function of age provide a quantitative illustration of these intense changes (adapted from (Dubois et al., 2019). b: The cortex also shows dramatic changes in microstructure complexity and maturation. A*

*recent multi-parametric MRI approach highlighted strong differences between cortical regions in infants between 1 and 5 months of age (coloured maps adapted from (J. Lebenberg et al., 2019)), in agreement with post-mortem map of subcortical myelination by (Flechsig, 1920) adapted from (Von Bonin, 1950)).*

### *2.2. Cortical morphology*

In parallel to volumetric changes, early brain development is characterized by dramatic changes in cortex morphology due to the cortical folding process that begins from 10 weeks of fetal life (Feess-Higgins & Larroche, 1987; Welker, 1988) (Figure 1a and Box 2). During the third trimester of pregnancy, the cerebral cortex changes from a relatively smooth, lissencephalic, surface to a complex folded structure that closely resembles the morphology of the adult cortex (Figure 1a). The precise mechanism underlying cortical folding is still unknown. However, several factors likely contribute to the prenatal processes that influence the shape of the folded cerebral cortex, including cortical growth, apoptosis (i.e., programmed cell death), differential expansion of superior and inferior cortical layers, differential growth of the cortical mantle relatively to the underlying white matter (Tallinen et al., 2016), transitory compartments such as the subplate (Rana et al., 2019), differential neuropil developments (Llinares-Benadero & Borrell, 2019) (Mangin et al., 2019), differential tangential expansion induced by a genetics-based protomap and/or structural connectivity through axonal tension forces (see (Borrell, 2018; Foubet, Trejo, & Toro, 2018; Kroenke & Bayly, 2018; Zilles, Palomero-Gallagher, & Amunts, 2013) for recent reviews on the phylogenetic, cellular and mechanical factors of the cortical folding process). The folding process might enable to increase the cortical surface area while maintaining the volume of the axons required for interconnecting the areas and thus relatively short distances and reasonable communication times between distant brain regions (Klyachko & Stevens, 2003). Macroscopic (morphological/volumetric) and microscopic (cellular) features of the cortex are therefore intrinsically interrelated. For instance, it has been shown that the cortical ribbon is thicker in the gyral areas and thinner in the sulcal areas and neurons located in the deep layers of gyri are squeezed from the sides and appear elongated while neurons that reside in the deep layers of sulci are stretched and look flattened (Hilgetag & Barbas, 2006, 2009).

Dedicated MRI tools and morphometric analyses have enabled to map in detail the developing cortical surface and growth patterns in fetuses as young as 20 wGA (Habas et al., 2012). These *in vivo* studies confirm earlier *post-mortem* observations (Chi, Dooling, & Gilles, 1977) and show a precise timing : stable primary folds appear around 20 wGA, secondary folds around 32 wGA and highly variable tertiary folds around term (Chi et al., 1977). The gyrification (the apparition of the gyri, the ‘mountains’ of the cortical relief) and the sulcation (the apparition of the sulci, the ‘valleys’ of the cortical relief) become manifest after 24 wGA (Rajagopalan et al., 2011), and greatly heightens during the last weeks before birth (Angleys et al., 2014; Dubois, Benders, Cachia, et al., 2008; Dubois et al., 2019). Although some variability is observed among individuals, the regional pattern is consistent over the brain surface: sulcation starts in the central region and proceeds first towards the parietal, temporal and occipital lobes, second towards the frontal lobe (Dubois, Benders, Cachia, et al., 2008; Ruoss, Lovblad, Schroth, Moessinger, & Fusch, 2001). The heritability of the cortical folding is estimated between 0.2 and 0.5 (Le Guen et al., 2018), supporting a major role of early environmental factors like alcohol exposure (De Guio et al., 2014), intrauterine growth restriction or twin pregnancy (Dubois, Benders, Borradori-Tolsa, et al., 2008) in the cortical folding process. The morphology of the cortical folding at birth can then be considered a retrospective marker of fetal brain development. It can also be used as prognostic marker of later functional development. Indeed, cortical sulcation at birth in preterms has been shown to predict infants' neurobehavioural development several weeks later (Dubois, Benders, Borradori-Tolsa, et al., 2008; Kersbergen et al., 2016).

The development of the volumetric and morphological features of the brain, although governed by specific mechanisms, are not independent. The cortical folding index scales uniformly across species and individuals as a function of the product of CSA and the square root of CT (Mota & Herculano-Houzel, 2015). As early as 23 wGA, CSA and total brain volume were also found to be related by a scaling law whose exponent predicts later neurodevelopmental impairment in preterms (Kapellou et al., 2006).

At birth, the cortical surface area is three times smaller than in adults, but the cortex is roughly similarly folded, and the most variable regions among individuals are the same across newborns and adults (Hill et al., 2010). Unlike quantitative features of cortical anatomy such as CSA or CT which can take decades to attain the levels observed in adulthood (Giedd & Rapoport, 2010; Li et al., 2014; Raznahan et al., 2011), the qualitative features of the cortex anatomy, such as the sulcal patterns (Cachia et al., 2016; Tissier et al., 2018) or the incomplete hippocampal inversion (Bajic et al., 2008; Cury et al., 2015) are determined during in utero and are stable after birth. The analysis of such trait features of the brain can thus provide information on the prenatal constraints imposed by the structure of some specific brain regions on later cognitive development.

### 2.3. Study of the fetal foundation of cognition using the sulcal patterns

Several studies have reported that subtle variations of the *in utero* environment, as indexed by birth weight, are accompanied by differences in postnatal cognitive abilities (Raznahan, Greenstein, Lee, Clasen, & Giedd, 2012; Shenkin, Starr, & Deary, 2004; Walhovd et al., 2012). In addition to such a global proxy measure of "uterine optimality" (Raznahan et al., 2012), analysis of the sulcal patterns can provide information on the prenatal constraints imposed by the structure of some specific brain regions on later cognitive development.

Several studies in typically developed participants reported long-term influence of cortical sulcation at birth on cognition several years and decades later (for a review in impaired cognition, see (Mangin, Jouvent, & Cachia, 2010)). For instance, a critical region of the cognitive control is the dorsal anterior cingulate cortex (ACC) (Petersen & Posner, 2012) which presents two qualitatively distinct sulcal patterns: a 'single' type when only the cingulate sulcus is present and a 'double parallel' type when a paracingulate sulcus runs parallel to the cingulate sulcus. An asymmetrical sulcal pattern of the ACC (i.e. different sulcal pattern in left and right hemispheres) was found to be associated with higher cognitive control in children at age 5 (Cachia et al., 2014), and 4 years later (G. Borst et al., 2014), as well as in adults (Fornito et al., 2004; Tissier et al., 2018). A similar effect was found in children and adults for the inferior frontal cortex (IFC), another key region of cognitive control (Tissier et al., 2018). These early neurodevelopmental constraints on later cognitive efficacy are not fixed nor deterministic. Indeed, only a part (15-20%) of the cognitive variability is explained by the sulcal pattern variability (G. Borst et al., 2014; Cachia et al., 2014; Tissier et al., 2018). In addition, different environmental backgrounds, either after birth such as bilingualism (Cachia, Del Maschio, et al., 2017; Del Maschio et al., 2018) or before birth such as twin pregnancy (Amiez, Wilson, & Procyk, 2018), can modulate the effect of the sulcal pattern on cognition. Sulcal studies also revealed that cognitive abilities requiring intensive learning and training, such as numeracy or literacy, can also be traced back to early stages of brain development. Indeed, the pattern (continuous or interrupted sulcus) of the posterior part of the left lateral occipito-temporal sulcus (OTS) hosting the visual word form area (VWFA) predicts reading skills in 10-years-old children (G Borst et al., 2016) and in adults (Cachia et al., 2018). The presence or absence of branches sectioning the horizontal branch of the intra-parietal sulcus (HIPS), a key region for processing numbers, is related to individual differences in symbolic number comparison and math fluency abilities in children and adults (Roell et al., Submitted).

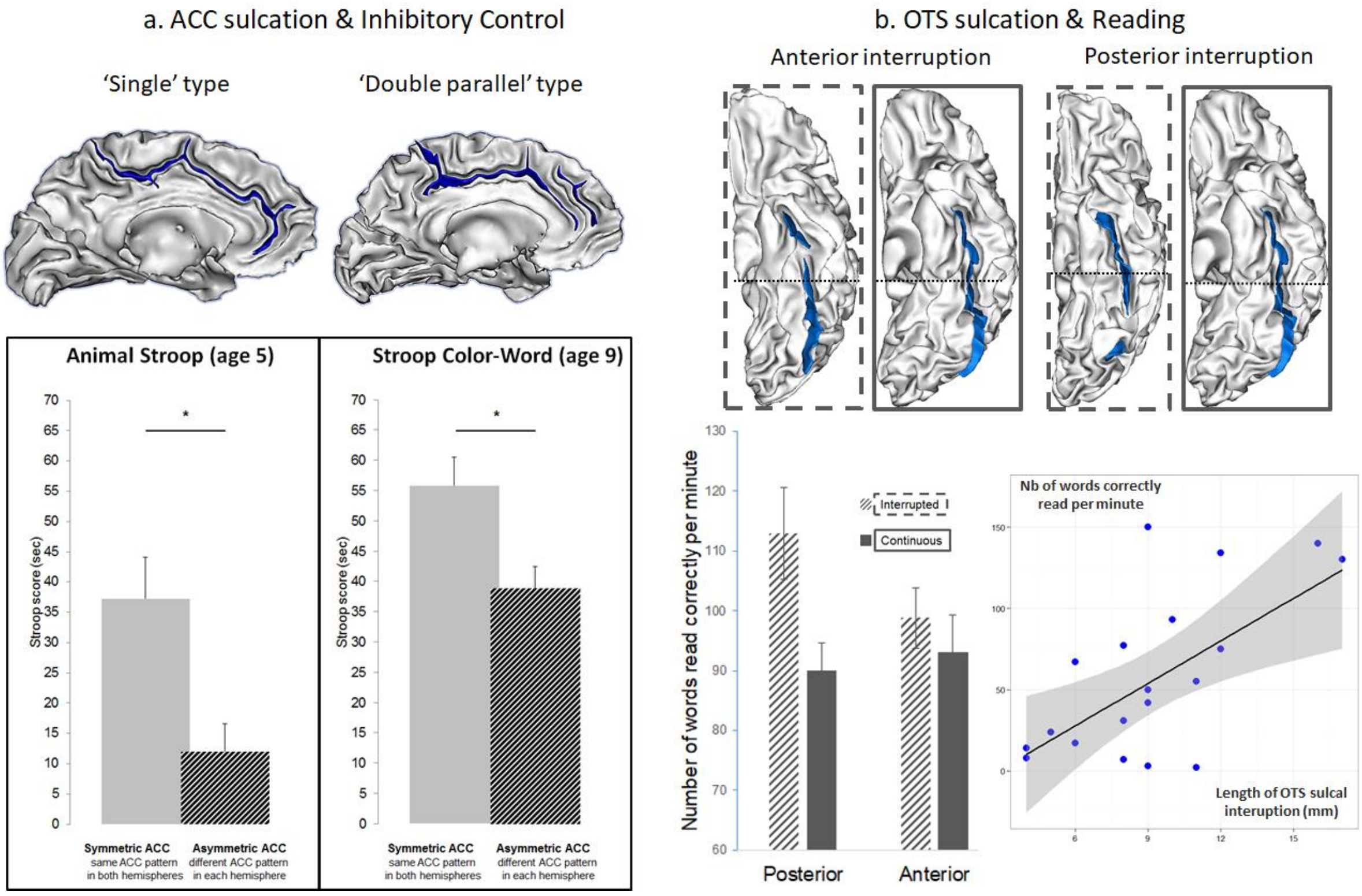

***Figure 2: Cortical sulcation and cognitive efficiency.*** *a. Effects of the sulcation of the anterior cingulate cortex (ACC) on inhibitory control efficiency. Upper panel: the ACC can have two types of sulcal patterns: 'single' type when only the cingulate sulcus was present and 'double parallel' type when a paracingulate sulcus ran parallel to the cingulate sulcus (sulci are depicted in blue on the gray/white interface). Lower panel: Mean Stroop interference scores at age 5 (on the Animal Stroop task) and at age 9 (on the Color-Word Stroop task) in children with symmetrical (single or double parallel type in both hemispheres) and asymmetrical (single type in the right hemisphere and double type in the left hemisphere or vice versa) ACC sulcal patterns. Children with asymmetrical ACC sulcal patterns have better inhibitory control efficiency than children with symmetrical ACC sulcal patterns. Adapted from (G. Borst et al., 2014; Cachia et al., 2014) b. Effect of Occipito-Temporal Sulcus (OTS) sulcation on reading ability. Upper panel: the left*

*posterior OTS hosting the visual word form area (VWFA) can have two types of sulcal patterns: 'continuous' type or 'interrupted' type in case of a sulcal interruption located posteriorly, below the back dashed line. Lower panel: Number of words reads correctly in children or adults participants with continuous (in plain gray) or interrupted (in hatched gray) left OTS. Participants with interrupted OTS have better reading efficiency than participants with continuous OTS. The number of words read per minute is positively correlated with the length of the sulcal interruption. Adapted from (G Borst et al., 2016; Cachia, Roell, et al., 2017).*

### *2.4. Cortical microstructure*

These macro-structural changes, with increasing cortical folding and thickness with age, are probably the visible markers of the microstructural evolution occurring in the cortical plate over pre-term and early post-term, marked by synaptic outburst and pruning, modifications in dendritic branching and fiber myelination. For instance, histological studies showed that cortical synaptogenesis starts from 16w GA, and the peak in synapse density depends on the brain region (Huttenlocher & Dabholkar, 1997; Kwan, Sestan, & Anton, 2012).

In the recent years, several MRI methods have been used to provide quantitative information on these mechanisms *in vivo* (Box 3). Diffusion MRI, notably diffusion tensor imaging, or DTI (Huppi & Dubois, 2006), has shown the different brain layers (cortical plate, subplate, inner layer) in *post-mortem* fetuses from 13w GA (Huang et al., 2013; Huang et al., 2009) and *in vivo* preterm newborns from 25w PMA (Maas et al., 2004). In the cortical plate, a complex age-related evolution is observed for the diffusion properties of water molecules (Dudink et al., 2010; McKinstry et al., 2002). This tissue first shows some anisotropy from 26w GA/PMA with a radial orientation of the main tensor direction which might rely on the early radial deployment of glial fibers and apical dendrites of pyramidal neurons (Ball et al., 2013; Ouyang, Dubois, Yu, Mukherjee, & Huang, 2019). This transient radial alignment was quantified by a radiality index that measures the directional coherence between the diffusion tensor and the cortical surface (Eaton-Rosen et al., 2017). Thereafter the cortical plate becomes isotropic from around 36w GA/PMA with the elongation and complex branching of neuronal connections (basal dendrites for the pyramidal neurons and thalamo-cortical afferents), and the decrease in cortical anisotropy seems to stop at around 38w PMA (Batalle et al., 2018). Besides, some reported an increase in DTI mean diffusivity between 26w and 32w PMA (McKinstry et al., 2002), whereas most others have described a steady decrease with age (Batalle et al., 2018; Ouyang et al., 2019). This suggested competing mechanisms, such as decrease neuronal density associated with programmed cell death versus increase in glial cells, addition of neuropils between the neuronal somas, decreasing water content, etc. To decipher between these complex microstructural changes, recent studies have used more elaborate diffusion models (e.g. Neurite Orientation Dispersion and Density Imaging, NODDI) (Batalle et al., 2018; Eaton-Rosen et al., 2015). Before 38w PMA, the neurite density index (NDI) decreases and the orientation dispersion index (ODI) increases, consistently with a predominant increase in dendritic arborisation and neurite growth. After this age, ODI plateaus and NDI increases in primary sensori-motor regions, suggesting that the 38-47w PMA period might be dominated by increasing cellular and organelle density. Thus these microstructural changes are not linear at least for some regions, and they are not uniform over the brain (Ball et al., 2013; Deipolyi et al., 2005). In particular, cortical sulci might show a more complex microstructure than gyri early on (Ball et al., 2013).

Diffusion parameters are also sensitive to the regional heterogeneity in cortical development. The occipital lobe shows the most rapid decrease in radiality index, and the frontal and temporal the least ones (Eaton-Rosen et al., 2017). Similarly, functionally distinct regions show different patterns of anistropy decrease from 20 to 35w GA (Yu et al., 2017), which highlights the asynchronous rate of maturation across cortical areas. DTI parameters further revealed advanced cortical maturation of the primary auditory cortex by 28w PMA, while rapid changes take place in the non-primary cortex of Heschl's gyrus between 26w and 42w PMA (Monson et al., 2018). In a recent exploratory study, sharper changes in cortical microstructure were related to a more rapid increase in cerebral blood flow in preterms from 32w to 45w PMA (Ouyang, Liu, et al., 2017).

During infancy, the evolution of cortical microstructure is complex. The growth of connections between neurons is first intense and exuberant through synaptogenesis and the extension of dendritic arborization. Then this phase is followed by an elimination of useless connections through a pruning mechanism, to select and maintain only functionally efficient connections. This occurs over different age periods for different cortical regions depending on the functions and the environmental stimulations. Accompanying this process, the dendritic and axonal fibers get myelinated, mostly during the early post-term period. *Post-mortem* studies showed that brain regions have different maturational trajectories relatively to the myelination of sub-cortical white matter fibers (Flechsig, 1920). This mechanism can be explored *in vivo* with MRI by taking advantages of changes in T1 and T2 signals (two complementary standard macroscopic measures depending on different features of the brain anatomy at cellular level) induced by modifications in water and iron contents (Ouyang et al., 2019). Differences in myelination across cortical regions get stronger from 36w to 44w PMA as observed in premature infants with the T1w/T2w ratio (Bozek et al., 2018). The asynchrony of maturation across primary and associative regions was also recovered over the infant language network during the first post-natal months based on T2w images (Leroy et al., 2011), and during the second year based on T1w images (Travis et al., 2014). Multiparametric MRI approaches combining DTI, relaxometry (e.g. quantitative T1) or multi-compartment approaches (e.g. myelin water fraction) have also highlighted differences in the microstructural and maturational properties of cortical regions in preterms at term equivalent age (Friedrichs-Maeder et al., 2017), in infants (J. Lebenberg et al., 2019) (Figure 1b) and toddlers (Deoni, Dean, Remer, Dirks, & O'Muircheartaigh, 2015). All these changes observed with distinct MRI measurements probably reflect a complex interplay between several mechanisms, such as the development of dendritic arborization, the proliferation of glial cells, and/or the myelination of intra-cortical fibers.

### 2.5. Maturation of central grey matter nuclei

Marked microstructural changes are also observed in central gray nuclei throughout development, with age-related decrease in DTI diffusivities and increase in anisotropy over the preterm period and infancy (Mukherjee et al., 2001; J. J. Neil et al., 1998; Qiu et al., 2013), suggesting intense membrane proliferation and fiber myelination. In parallel and simultaneously with white matter regions, T1 decreases gradually over central gray nuclei (Schneider et al., 2016), as T2 does (Bultmann, Spineli, Hartmann, & Lanfermann, 2018). The developing microstructural properties are better characterized by comparing multiple parameters (e.g. DTI, NODDI, T1, magnetization transfer ratio, fraction of water related to myelin). This enable to disentangle between maturational mechanisms (e.g. concentration of myelin-associated macromolecules, water content) (Melbourne et al., 2016; Nossin-Manor et al., 2013). Much like the cortical regions, the maturational patterns differ across nuclei, notably across thalamic substructures (Poh et al., 2015).

## 3. The early development of white matter

In interaction with the development of cortical regions, intense and intermingled processes of growth and maturation are occurring within the brain white matter from the preterm period to infancy (Dubois, Adibpour, Poupon, Hertz-Pannier, & Dehaene-Lambertz, 2016; Dubois et al., 2014; Dubois et al., 2015). All major long-distance fibers are observed by term birth (Figure 3a), while short-distance fibers (e.g. U-fibers) mainly develop during the first post-natal year. All these anatomical connections further refine through several complementary mechanisms. After an exuberant growth and proliferation, useless and redundant connections are pruned during childhood, whereas the ongoing myelination process stabilizes the functionally relevant ones and increases the efficiency of the information transfer between distant brain regions. Whereas the number of neurons and microglia remains almost the same post-natally, the number of oligodendrocytes and astrocytes drastically increases in the white matter during the first 3 years, attaining two-thirds of the corresponding numbers in adults (Sigaard, Kjaer, & Pakkenberg, 2016). The inhibitory role of oligodendrocytes and myelin on neuritic growth may partly explain the weak plasticity of the adult brain (Ng, Cartel, Roder, Roach, & Lozano, 1996). In addition to the early brain organization in specific networks, a major developmental characteristic is the asynchronous progression of maturation across brain regions: for instance, sensory regions develop early on and quickly, whereas associative regions are slowly developing until the end of adolescence.

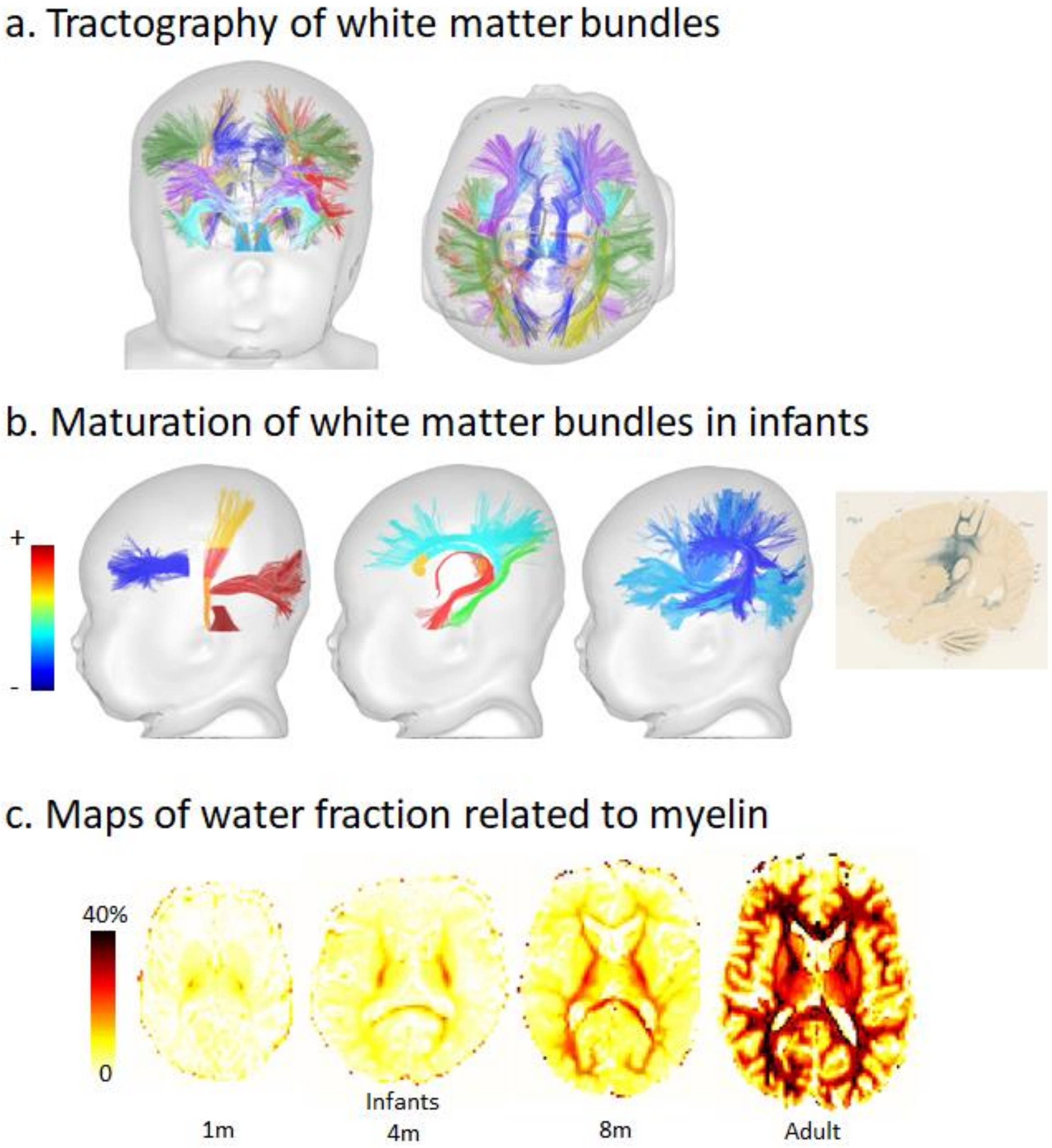

***Figure 3: Structural changes of the white matter during development.*** *a: The brain shows an early organization in networks, with white matter bundles connecting distant and close regions, as illustrated here with diffusion MRI and tractography reconstructions in a 1-month old infant. b: The white matter also shows intense maturation after birth through the myelination process (P. I. a. L. Yakovlev, A. R., 1967). A recent multi-parametric MRI approach confirmed strong differences between bundles in infants between 1 and 5 months of age (adapted from (Dubois et al., 2015; Kulikova et al., 2015) in agreement with post-mortem map of subcortical myelination by (Flechsig, 1920) adapted from (Von Bonin, 1950)). c: Different MRI techniques inform on myelination, such as the one computing the*

*fraction of water related to myelin. These maps highlight the progression of myelination from central regions to the periphery (adapted from (Kulikova, Hertz-Pannier, Dehaene-Lambertz, Poupon, & Dubois, 2016)).*

### 3.1. Developing architecture of white matter bundles

Recent *post-mortem* and *in vivo* studies using diffusion MRI combined with physical models (e.g. diffusion tensor imaging DTI, high angular resolution diffusion imaging HARDI) and sophisticated tractography tools (cf. Box 3) have confirmed and extended the knowledge on bundles growth and organization which previously relied on histological dissections. Nevertheless, several issues are related to these techniques in the developing brain (Dubois, Adibpour, et al., 2016). For instance, fetal and neonatal MRI is limited by motion-related constraints and spatial resolution issues (Dubois et al., 2014). Besides, the biological interpretation is not straightforward: not only the axons participate to the depicted anisotropy of water diffusion, but also the radial glial fibers and the penetrating blood vessels, particularly during the preterm period (Xu et al., 2012).

The development of fiber systems proceeds in a sequential way (projection, callosal, and association) (Vasung, Raguz, Kostovic, & Takahashi, 2017), and transient fetal patterns of connectivity are observed before the formation of stable connections (Dubois et al., 2015; Kostovic & Jovanov-Milosevic, 2006; Vasung et al., 2010). During the early fetal period (9-15 weeks post-conception wPC), major fiber pathways are growing and pathfinding within the intermediate zone. Some afferent (e.g. thalamo-cortical) and efferent (cortico-subcortical) fibers are already observed, as well as several limbic bundles (e.g. fornix, cingulum) (Huang et al., 2006; Vasung et al., 2010). Before penetrating into the cortical plate, thalamo-cortical fibers make connections with neurons of the subplate zone (Kostovic & Rakic, 1990). During the midfetal period (15-23 wPC), major efferent fibers penetrate their targets in sub-cortical structures, and cortico-cortical callosal fibers and some associative connections (e.g. uncinate, inferior longitudinal, fronto-occipital fasciculi) begin to emerge (Huang et al., 2009; Judas et al., 2005; Vasung et al., 2017). In fetuses *in utero*, the pyramidal tract, the splenium and genu of the corpus callosum are observed from 18 weeks of gestational age (wGA) (Bui et al., 2006; Kasprian et al., 2008; Pontabry et al., 2013), while the uncinate and inferior fronto-occipital fascicles are depicted from 20w GA (Mitter, Prayer, Brugger, Weber, & Kasprian, 2015).

During the early preterm period (24-28 wPC), afferent fibers that were waiting in the subplate start making connections within neurons of the cortical plate (of the future cortical layer IV), leading to the establishment of the permanent connectivity while the transient fetal circuitry still exists within the subplate (Kostovic & Judas, 2006; Vasung et al., 2016). Limbic cortico-cortical connections are well developed in the cingulate, entorhinal and hippocampal cortices (Kostovic, Petanjek, & Judas, 1993). Some associative (e.g. inferior longitudinal fasciculus) and limbic bundles (cingulum, fornix) can be visualized in 3D in fetuses *in utero* (Mitter et al., 2015). During the late preterm period (29-34w PC), long associative and commissural bundles are quickly developing, likely originating from pyramidal neurons of cortical layer III (Schwartz & Goldman-Rakic, 1991), and these bundles can be identified in preterm newborns (Dudink et al., 2007; Partridge et al., 2004). Short cortico-cortical fibers further grow and enter the cortex through the remnant subplate (Kostovic, Kostovic-Srzentic, Benjak, Jovanov-Milosevic, & Rados, 2014; Takahashi, Folkerth, Galaburda, & Grant, 2012).

At term birth, elaborate connectivity is established, allowing infants to acquire functional abilities. Notably, in infants born preterm, cognitive performances at 2 years are related with thalamocortical connectivity at term-equivalent age (Ball et al., 2015). U-fibers further develop (Kostovic, Jovanov-Milosevic, et al., 2014), and cortico-cortical connectivity is reorganized by several processes including the pruning of callosal fibers (LaMantia & Rakic, 1990) which probably extends until the end of the first postnatal year (Kostovic, Jovanov-Milosevic, et al., 2014; Kostovic & Rakic, 1990; Petanjek et al., 2011). During early infancy, all major white matter bundles and short-range connections are identified despite their weak maturation: commissural bundles of the corpus callosum (genu, body and splenium), projection bundles (cortico-spinal and spino-thalamic tracts, optic radiations, anterior limb of the internal capsule), limbic bundles (fornix and cingulum) and associative bundles (external capsule, uncinate, arcuate, superior and inferior longitudinal fascicles) (Dubois, Hertz-Pannier, Dehaene-Lambertz, Cointepas, & Le Bihan, 2006; Kulikova et al., 2015). The trajectory and morphology of these bundles remain stable between birth and 2 years of age (Geng et al., 2012).

In the recent years, the developing architecture of anatomical networks has also been detailed based on connectivity matrices (Keunen, Counsell, & Benders, 2017), by measuring the degree of connections for all pairs of brain regions. Already at 30 weeks of post-menstrual age (w PMA), the structural connectome demonstrates a small-world modular organization like in the adult brain (M. P. van den Heuvel et al., 2015), and cortical hubs are highly connected to form a “rich club” (Ball et al., 2014). This topology further refines with age (M. P. van den Heuvel et al., 2015), becoming more clustered around term (Brown et al., 2014) and showing an increase in global efficiency and integration, and a decrease in segregation during the first two post-natal years (Yap et al., 2011).

### *3.2. White matter maturation*

Concurrently and subsequently to the development of connections, the white matter bundles progressively mature and become functionally efficient through the myelination process. It includes several steps, and proceeds from the neuron body to the periphery (McCart & Henry, 1994). It occurs in the human brain from the second part of pregnancy to the end of adolescence, with a peak during the first post-natal year (Baumann & Pham-Dinh, 2001; Van der Knaap & Valk, 1995a, 1995b). It progresses at different ages and rates depending on the regions, bundles and networks (Baumann & Pham-Dinh, 2001; Brody, Kinney, Kloman, & Gilles, 1987; Flechsig, 1920; Gilles, Shankle, & Dooling, 1983; Kinney, Brody, Kloman, & Gilles, 1988; P. L. Yakovlev, A. , 1967; Yakovlev, 1962): it follows a caudo-rostral gradient, from the center to the periphery, and it occurs earlier and faster in sensory pathways (somatosensory, vision, audition) than in motor ones, and in projection fibers than in associative ones (Figure 3b).

The myelination process can be quantified through several MRI parameters (Dubois, Adibpour, et al., 2016; Dubois et al., 2014), which show intense changes after term birth and during the first post-natal months, and differences in maturation across white matter bundles (Dubois, Dehaene-Lambertz, Perrin, et al., 2008; Dubois, Poupon, et al., 2016; Kulikova et al., 2015). With myelination, the water content decreases, thus the proton density decreases, while the macromolecule tissue volume (Mezer et al., 2013) increases (Yeatman, Wandell, & Mezer, 2014). This mechanism, together with changes in water molecules compartmentalization (Matsumae et al., 2001) and increase of protein and lipid contents (Barkovich, Kjos, Jackson, & Norman, 1988; Kucharczyk, Macdonald, Stanisz, & Henkelman, 1994), lead to decreases in T1 and T2 relaxation times with age during the “pre-myelinating” state and with the chemical maturation of the myelin sheath (Barkovich et al., 1988; Baumann & Pham-Dinh, 2001; Deoni, Dean, O'Muircheartaigh, Dirks, & Jerskey, 2012; Engelbrecht, Rassek, Preiss, Wald, & Modder, 1998; Haselgrove, Moore, Wang, Traipe, & Bilaniuk, 2000; Leppert et al., 2009; Poduslo & Jang, 1984).

DTI parameters also capture some aspects of this maturational pattern (Huppi & Dubois, 2006; J. Neil, Miller, Mukherjee, & Huppi, 2002). During the preterm period, diffusivities decrease, while anisotropy increases in most white matter regions (Kersbergen et al., 2014) except at cross-roads locations (Nossin-Manor, Card, Raybaud, Taylor, & Sled, 2015). During infancy and childhood, transverse diffusivity decreases more than longitudinal diffusivity (Dubois, Dehaene-Lambertz, Perrin, et al., 2008; Geng et al., 2012; Krogsrud et al., 2015), leading to anisotropy increase (Dubois et al., 2014). These parameters might be sensitive to the proliferation of glial cell bodies, the extension of oligodendroglial processes, and their ensheathment around axons (Dubois, Dehaene-Lambertz, Perrin, et al., 2008; Nossin-Manor et al., 2013; Nossin-Manor et al., 2015; Zanin et al., 2011). Several other MRI parameters can be measured to quantify the maturation as they vary with the density of myelin-associated macromolecules and axonal cytoskeleton components: magnetic susceptibility (Li et al., 2013), magnetization transfer imaging or ratio (Kucharczyk et al., 1994; McGowan, 1999; Xydis et al., 2006).

Recently, several modelling approaches have been proposed to characterize distinct pools of water molecules with different microstructural properties (Dubois, Adibpour, et al., 2016; Dubois et al., 2014). Based on relaxometry data, some have allowed to estimate the volume fraction of water related to myelin, which drastically increases with age in the white matter (Deoni et al., 2012; Deoni et al., 2011; Kulikova et al., 2016) (Figure 3c). Based on diffusion data, some have characterized the intra-neurite volume fraction that relies on the fibers maturation, and the neurite orientation dispersion that depends on fiber crossings and fanning (Dean et al., 2017; Kunz et al., 2014). Other approaches have taken benefit of several MRI parameters that provide complementary information in the developing white matter (Dubois, Adibpour, et al., 2016; Dubois et al., 2014). Larger axons and a thicker myelin sheath contribute to faster conduction, but there is a trade-off between axon size and myelin thickness due to spatial constraints imposed by the brain dimensions (Fields, 2008). The relationship between axon size and myelin thickness is captured in a parameter called the myelin g-ratio, defined as the ratio of the inner (axon) to the outer (axon plus myelin) diameter of the fiber. The g-ratio can be estimated using relaxometry and diffusion MRI data (Stikov et al., 2015). The g-ratio decreases with myelination (Dean et al., 2016; Melbourne et al., 2016) and its variations have been associated with cognition and disease (Fields, 2008). Finally, multi-parametric approaches have combined T1, T2 and DTI parameters with clustering (Jessica Lebenberg et al., 2015) or mathematical distance to the adult stage (Kulikova et al., 2015) to better highlight the patterns of maturation across regions and bundles.

### 3.3. *White matter maturation and functional development*

A few recent studies have related the developmental changes in white matter networks and the psychomotor acquisitions in infants and children. Based on the fraction of water related to myelin, cognitive scales for receptive and expressive language (O'Muircheartaigh et al., 2014), and processing speeds (Chevalier et al., 2015) have been linked to the maturation of fronto-temporal regions and of occipital region respectively. A longitudinal study over the first 5 years showed that children with above-average ability might have differential trajectories of maturation compared to average and below average ability children (Deoni et al., 2016). This suggested that infants might benefit from an early period of prolonged maturation associated with protracted plasticity, and independently of socioeconomic status, gestational age and weight at birth. The increasing accuracy and speed of processing complex sentences in children have been related to the DTI microstructural properties of the arcuate fasciculus which connects fronto-temporal regions of the language network (Skeide, Brauer, & Friederici, 2016). This tract anisotropy measured at term equivalent age in preterm-born infants is also associated with individual differences in linguistic and cognitive abilities at 2 years of age, independently of the degree of prematurity (Salvan et al., 2017). Considering structural connections over the whole brain, the white matter anisotropy in neonates has even been related to behavioral functioning at 5 years of age (Keunen, Benders, et al., 2017).

Despite informative, such behavioral measures correspond to complex processes that only reflect the white matter maturation in an indirect and composite way. In fact myelination is known to increase the conduction velocity of the nerve impulse along axonal fibers (Baumann & Pham-Dinh, 2001), which either leads to a decrease in the response latency at constant pathway length, or to a preserved latency compensating for a growing length (Salami, Itami, Tsumoto, & Kimura, 2003). Recently some studies tackled this issue by directly comparing structural and functional markers of the white matter maturation, combining MRI and electro- or magnetoencephalography (EEG/MEG) in the same subjects (Dubois, Adibpour, et al., 2016). These latter techniques enable to measure the latency of evoked responses (i.e. the averaged responses over multiple trials following successive stimulations), and to identify its age-related decrease in infants. The following studies investigated whether this gradual acceleration in the peak latencies during infancy might be used as a proxy of the bundles myelination (see (Dubois, Adibpour, et al., 2016) for a more extensive review).

The visual modality develops mainly after birth, and infants' visual capacities drastically improve in a few months. Successive EEG/MEG visual evoked responses are recorded in occipital regions. At term birth, the small EEG positive component P1 (~P100 in adults) is detected at a latency that decreases strongly and quickly with age (Harding, Grose, Wilton, & Bissenden, 1989; McCulloch & Skarf, 1991; Taylor, Menzies, MacMillan, & Whyte, 1987), from around 260ms in neonates to around 110-120ms at 12-14 weeks of age, depending on the patterns size (McCulloch, Orbach, & Skarf, 1999). This suggests progressive increase in conduction velocity in afferent visual pathways related to fiber myelination (Lee, Birtles, Wattam-Bell, Atkinson, & Braddick, 2012), together with other maturational processes (e.g. retina and cortical development).

These functional changes have been compared to the microstructural evolution of visual bundles throughout development, focusing on the latency of P1 and its transfer from contralateral to ipsilateral hemisphere in case of lateral stimuli. While taking the infants' age into account, the variability in P1 conduction velocity across infants has been related to the DTI maturation of optic radiations (P. Adibpour, Dubois, & Dehaene-Lambertz, 2018; Dubois, Dehaene-Lambertz, Soares, et al., 2008) (Figure 4a). This observation has been further extended to cortico-cortical connections. When stimuli are presented laterally (i.e. in a single hemifield), visual responses are first observed in the contralateral hemisphere, then in the ipsilateral hemisphere. In infants, the velocity of this inter-hemispheric transfer has been related to the maturation of the visual fibers of the corpus callosum (P. Adibpour, Dubois, & Dehaene-Lambertz, 2018) (Figure 4b), in a similar way as P1 and optic radiations. In 6-12-year-old children, the MEG P1 latency has been further related to the white matter properties of visual and motor association regions (Dockstader, Gaetz, Rockel, & Mabbott, 2012).

So far, no developmental study has related the anatomical and functional maturation in the somatosensory modality, despite intense development during the preterm period and major changes in electrophysiological responses throughout infancy (Dubois, Adibpour, et al., 2016). Only a controversial study in adults (Horowitz et al., 2015; Innocenti, Caminiti, & Aboitiz, 2015) has suggested that the N140 inter-hemispheric transfer time depends on the axon diameter of callosal fibers connecting primary somatosensory cortices.

As for the auditory modality, it is already functional *in utero*, but its development is more protracted throughout infancy and toddlerhood than the visual and somatosensory modalities. Thus evoked responses show extended developmental changes throughout the first post-natal years (Dubois, Adibpour, et al., 2016), and this might make the comparison with white matter maturation difficult in infants (P. Adibpour, Dehaene-Lambertz, & Dubois, 2015).

Following auditory monaural stimulations (i.e. syllables presented in one ear at a time), early responses (P2) are observed in infants both on the contralateral and ipsilateral sides because, as in the adult brain, both types of pathways are already functional (P. Adibpour, Lebenberg, Kabdebon, Dehaene-Lambertz, & Dubois, 2020; P. Adibpour, Dubois, Moutard, & Dehaene-Lambertz, 2018). The latencies of these different responses decrease with age (P. Adibpour et al., 2020), and ipsilateral responses are significantly longer in the left than in the right hemisphere, which is not the case in infants with agenesis of the corpus callosum (P. Adibpour, Dubois, Moutard, et al., 2018). These results might suggest that contralateral responses are transmitted, via callosal fibers, more from the right to the left hemisphere than the opposite. Such functional asymmetries could influence the early lateralization of the language network and reinforce an initial bias during development. The comparison of DTI, structural and EEG functional measures in infants further outlined that the speed of the left ipsilateral response (the same one that is slower in the right hemisphere) tends to vary with the microstructure of auditory callosal fibers (namely the more these fibers have an advanced maturation, the higher the speed (P. Adibpour et al., 2020)) (Figure 4c), suggesting the involvement of auditory callosal fibers in the emergence of this lateralized response. Afterwards, the decrease in the latency of MEG P2m during childhood has been related to the DTI maturation of acoustic radiations (Roberts et al., 2009).

Reliable correlations between the properties of evoked responses and the maturation of white matter pathways remain scarce so far. Further studies in the somatosensory and auditory modalities are required to confirm the results obtained in the visual modality. And comparing the anatomo-functional changes across modalities within the same subjects would help to characterize the asynchronous development of brain networks and explore their sensitivity to distinct critical periods and to various environmental stimulations.

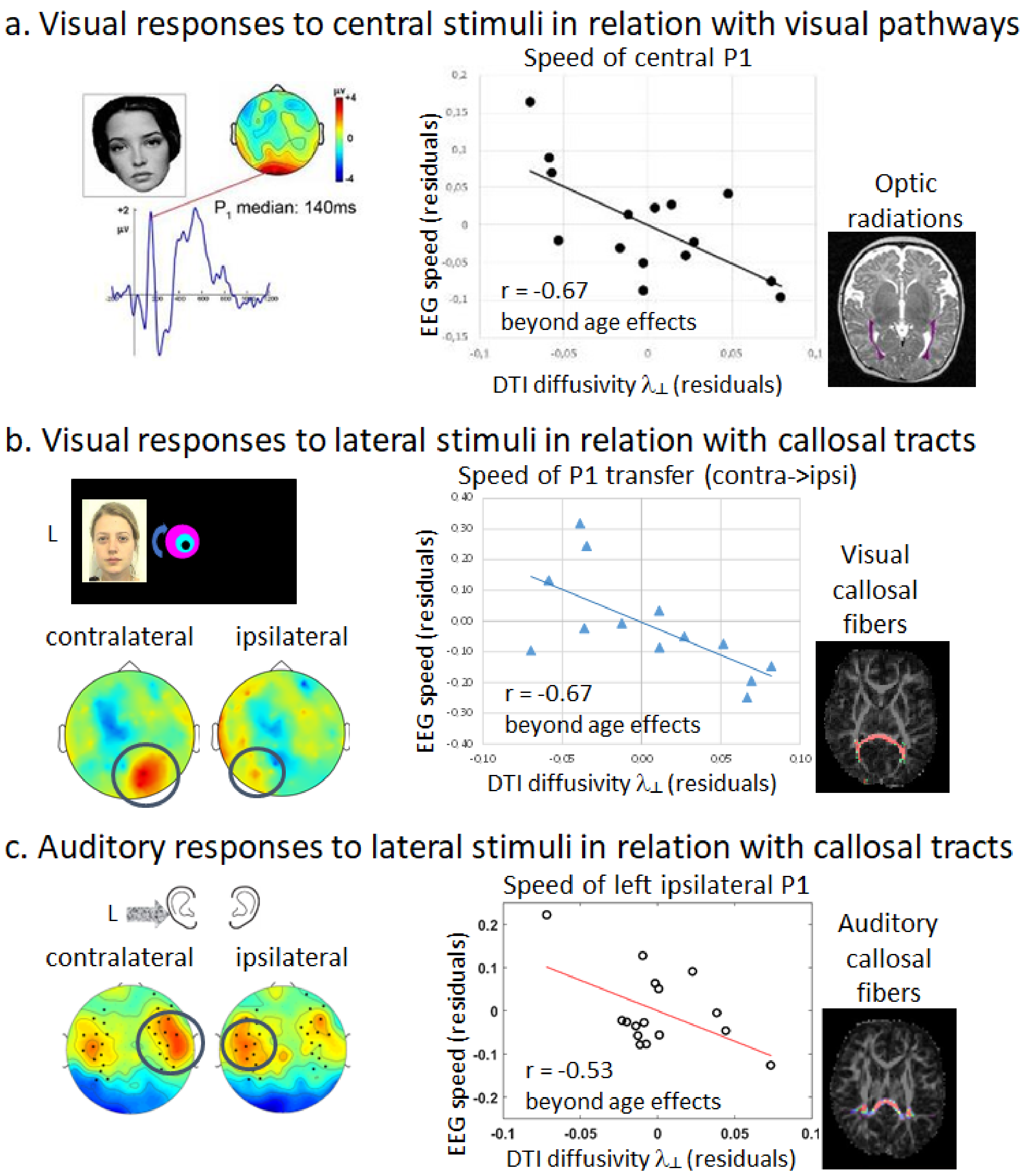

***Figure 4: Relationships between functional and structural markers of development.*** *a: During infancy, the speed of P1 responses to visual stimuli increases with age, while the visual pathways become myelinated, resulting in changes of DTI indices (e.g. decrease in transverse diffusivity l⊥ in the optic radiations). Recent studies have shown that these functional and structural markers of visual maturation are related beyond age dependancies (adapted from (Dubois, Dehaene-Lambertz, Soares, et al., 2008). b: Similar relationships have been observed for responses to visual stimuli presented laterally (in one hemifield at a time). The speed of the responses transfer, from the contralateral to the*

*ipsilateral hemisphere, is related to the maturation of callosal fibers connecting visual regions (adapted from (P. Adibpour, Dubois, & Dehaene-Lambertz, 2018)). c: Such relationships are more difficult to demonstrate for the auditory system because of strong asymmetries in the latency of responses to lateral stimuli (presented in one ear at a time) (P. Adibpour, Dubois, Moutard, et al., 2018). It seems that the speed of the left ipsilateral responses is related to the maturation of callosal fibers connecting auditory regions (adapted from (P. Adibpour et al., 2020). This suggests that early structural biases might lead to the functional lateralization for speech processing in the left hemisphere.*

# 4. Perspectives: toward an integrative approach

Besides further investigations of each piece of the complex puzzle of the early brain development, a critical challenge consists in combining all the pieces together so as to provide an integrative view of the neurocognitive development. This integrative approach raises methodological issues to characterize processes spanning different structures, scales, functions, and ages (Betzel & Bassett, 2017; Martijn P van den Heuvel, Scholtens, & Kahn, 2019) and require sound epistemological debates for interdisciplinary cross-talks. We provide below recent integrative attempts regarding: i) connectivity development at the structural and functional levels and ii) white matter connectivity and grey matter maturation.

## *4.1. Relating the development of the structural and functional connectivity*

Recent advances in neuroimaging techniques have enabled the simultaneous quantification of both anatomical and functional connectivity at the macroscale in pediatric populations (Ouyang, Kang, Detre, Roberts, & Huang, 2017). These studies have reported hierarchical structural maturation from primary to higher-order cortices, with important individual variability, which is partially paralleled by functional development (Cao, Huang, & He, 2017). For instance, the thalamo-cortical connectivity patterns in infants reveals good spatial agreement between structural and functional MRI modalities in general but with regional variations that are system-specific (Ferradal et al., 2018) : regions involving primary-sensory cortices exhibit greater structural/functional overlap, whereas higher-order association areas such as temporal and posterior parietal cortices show divergence in spatial patterns of each modality. In addition, analysis of the structural and functional brain networks from fetal life to 2 years of age suggest that the structural network remains ahead and paves the way for the development of the functional brain network (Zhao, Xu, & He, 2018) and cognitive and behavior performance (Wee et al., 2017). The development of the functional connectome is reconfigured through the integration and segregation processes, contributed by increasing long-range functional connectivity and decreasing short-range functional connectivity respectively (Cohen et al., 2008; Fair et al., 2009; Fair et al., 2007). Synaptic over-growth and the pruning that follows it as well as the myelination process, regulated by spatiotemporal transcriptomic profiles (Silbereis, Pochareddy, Zhu, Li, & Sestan, 2016), are the major factors for reshaping the development and tuning of functional brain networks. The bridge between these microscale neuronal processes and macroscale connectome reconfiguration still need to be directly established because current in vivo non-invasive neuroimaging techniques have low spatial resolution (around 1 mm) that cannot approach cellular scales.

### *4.2. Relating the patterns of grey and white matter maturation*

Spatial and functional gradients of development have been described for the dynamic interplay of cerebral gray and white matter using histological and multimodal neuroimaging approaches. Synchronous development between cortical gray matter (GM) and adjacent white matter (WM) were observed in primary motor, primary visual, visual association, and prefrontal regions, with maturation in primary motor and sensory regions preceding maturation of association areas (Smyser et al., 2016). Around term age, the development of thalamic substructures - connecting to the frontal, precentral, postcentral, temporal and parieto-occipital cortices - is synchronized with the maturation of their respective thalamo-cortical connectivity (Poh et al., 2015). More generally, the advancement of GM and WM maturation is inter-related and dependent on the underlying brain connectivity architecture (Friedrichs-Maeder et al., 2017). Particularly, corresponding maturation levels are found in GM regions and their incident WM connections, and also in GM regions connected through a WM tract. From 1 to 6 years of age, regional measures of cortical thickness were found to be partially driven by changes in adjacent white matter myelination, suggesting that cortical and white matter maturation reflect distinct, but complimentary, neurodevelopmental processes (Croteau-Chonka et al., 2016).

## 5. Conclusion

Human brain development is a complex and dynamic process that begins during the first weeks of pregnancy and lasts until early adulthood. This chapter focused on the developmental window from prenatal period to infancy and showed how the cortex grows and gets convoluted, the microstructural maturation of the gray and white matter and their cognitive correlates in normal condition. To summarize, the last weeks of pregnancy and the first postnatal months are marked by an intense increase in cortical volume which progressively slows down after 2 years of age until adolescence. These changes are driven by changes in cortical thickness and cortical surface area, two complementary macrostructural features of the cortex anatomy with distinct genetic and developmental mechanisms. In parallel to volumetric changes, early brain development is characterized by dramatic change in cortex morphology due to the cortical folding process that begins from 10 weeks of fetal life. During the third trimester of pregnancy, the cerebral cortex changes from a relatively smooth surface to a complex folded structure that closely resembles the morphology of the adult cortex. The morphology of the cortical folding is a marker of early neurodevelopment predisposition and its analysis has evidenced the long-term effect of fetal life on later cognitive development. These macro-structural changes of the cortex are associated with cellular changes. The growth of neuronal connections is first intense through synaptogenesis and the extension of dendritic arborization and then followed by an elimination of useless connections through a pruning mechanism, to select and maintain only functionally efficient connections. This occurs over different age periods for different cortical regions depending on the functions and the environmental stimulations. In interaction with the cortical development, intense and intermingled processes of growth and maturation are also occurring within the white matter. At birth, elaborate connectivity is established, allowing infants to acquire functional abilities. Concurrently and subsequently to the development of connections, the white matter bundles progressively mature and become functionally efficient through the myelination process, early and fast for the sensory regions while slow and long-lasting for the associative regions. All these cellular and macroscopic changes in white matter networks contribute to cognitive development in infants and children, but direct evidence of such relationships remains scarce so far. Finally, besides further investigations of these different age-, scale- and function-specific processes, the ultimate challenge will be to combine all these complementary facets of the neurocognitive development into a comprehensive integrative model.

# 6. BOX

## *6.1. BOX 1: Measuring the cerebral volumes*

The first approaches to quantifying the volume of a brain structure from an MRI image were simply hand-drawing them. Gradually, this tedious approach was replaced by algorithmic methods for automatically segmenting regions of interest (ROI). There is a wide variety of such methods, but the most common ones are based on software freely distributed throughout the community (Freesurfer, FSL, SPM, volbrain, etc.). Most of them fit a brain template to the brain to be segmented by deforming it. This brain template is associated with an atlas of predefined ROI, generally hand-drawn, which then automatically adjusts to the structures of the brain to be segmented. Some particularly complex structures such as the hippocampus and its subfields require dedicated developments. The cortex itself, due to its surface geometry, often requires considering not only the volume of one of its subdivisions, but also its average thickness (Fischl, 2012). During the last decade, it has become important to use for segmentation not a single atlas but a set of atlases in order to benefit from a better modelling of the inter-individual anatomical variability (Manjón & Coupé, 2016). For each brain structure, the atlas that most closely resembles the brain to be segmented is used as the model. Usual automatic segmentation methods frequently require hours of computation to segment a brain, but deep-learning approaches are on the way to performing the analysis in a few seconds.

A notorious weakness of the ROI approach is that it imposes a parcellation of the brain which is not necessarily ideal for the question asked. If the anatomical structure of interest does not behave homogeneously, for instance relative to its maturation or its involvement in a physio-pathological process, estimating its overall volume does not provide much sensitivity. To overcome this difficulty, the neuroimaging community uses a strategy that comes from the world of neurosurgical planning, which has become an inescapable paradigm for functional mapping. It provides brains with a universal coordinate system that matches them without explicit concern for the details of their anatomy. This coordinate system is the result of an algorithmic operation very similar to the one used for automatic segmentation: the brain to be studied is deformed to make it look as much as possible like a template brain endowed with the coordinate system. This operation is called ‘spatial normalization’. The coordinate system then plays the role of a Global Positioning System (GPS). It should be noted that this operation also aligns the structures of this brain with those of the template brain, but this is not really used for the coordinate-based analyses. It is then possible to quantify the brain volume locally, point by point, without any a priori hypothesis (Ashburner & Friston, 2000). The most frequent brain template is an average brain obtained from a large number of previously normalized brains. There is an iterative side to this construction of the average brain which is regularly refined by improving brain alignment techniques. The ancestral three-dimensional coordinate system coming from neurosurgery is called the Talairach space. It corresponds to a single brain used by neurosurgeons as an atlas. Today we more commonly use the MNI space, proposed by the Montreal Neurological Institute, derived from the Talairach space but based on an average of 152 brains. When the object of study is the cortex, one often prefers to opt for a system with only two coordinates very similar to the longitude and latitude used to locate oneself on the surface of the earth (Fischl, 2012).

What do you measure once a brain has been spatially normalized? The simplest possibility is to focus on a tissue, the grey or white matter, and quantify the amount of tissue aggregated in each voxel of the image (a 3D pixel) after spatial normalization. This quantity can be deduced from the deformation field allowing the studied brain to be adjusted to the brain template. Usually, we do not take into account the global scaling factors that allow at the beginning of the normalization process to correct global differences in length, width or height with respect to the template. This strategy is called Voxel-Based Morphometry (VBM). An alternative is to compare the deformation fields themselves, which can shed light on the nature of the spatial transformations that result, for example, from brain development or an atrophy process. This is called Deformation-Based Morphometry (DBM). When focusing on the cortex using a two-dimensional coordinate system, the thickness of the cortex at each point on its surface is compared across subjects. It should be noted that the current consensus is to extract this thickness in the native space of the subject, and thus to discard any correction linked to the dimensions of the brain.

The different brain morphometric strategies mentioned above are now relatively stabilized and applied systematically throughout the community. On the other hand, when it comes to studying brain development, many methodological questions remain relatively open. With regard to the ROI approach, automatic segmentation methods are less well developed, especially as one moves backwards in age, but the rise of deep learning seems to be solving the problem. However, it will still have to be taken into account that as a structure matures, MRI contrasts change and do not necessarily allow volumes to be quantified in an equivalent manner from one age to another. For approaches based on spatial normalization, the issue is more complex, as it is difficult to free normalization from the effects of development. The first difficulty is the same as for ROI approaches, the evolution of MRI contrasts with maturation impacting alignment with the brain template. But the major difficulty lies in the choice of the brain template. Using an adult brain template has consequences that are difficult to understand on the final interpretation of the results. Using a family of templates dedicated to each age group is often preferred, but then it remains to align the templates of this family as well as possible. This question becomes particularly difficult when looking at the antenatal period where the patterns of cortical folding differ significantly from one age group to the next. An interesting strategy is to use alignment methods that explicitly impose the alignment of specific cortical sulci (J. Lebenberg et al., 2018).

### *6.2. BOX2: Measuring the cortex morphology*

Despite the maturity of morphometric approaches based on spatial normalization, some questions tend to resist them with respect to cortex morphology. Indeed, the inter-individual variability of cortical sulci is such (Figure 5) that its consequences on normalization are not really understood (Mangin et al., 2016). Consequently, detecting a difference between two groups of individuals with VBM does not necessarily imply a difference in volume but sometimes essentially a local difference in the folding pattern. The strategy based on the measurement of cortex thickness does not escape this difficulty, especially since there is a close correlation between cortical surface geometry and cortex thickness: the cortex is noticeably thicker at the top of the gyri than at the bottom of the folds (Wagstyl et al., 2018).

The first studies of cortical morphology were based on the concept of the gyrification index (GI), a ratio that makes it possible to evaluate the amount of cortex buried in the folds. Historically, this ratio was based on contour lengths measured in 2D sections, which led to major biases. Today, thanks to the automatic segmentation of the cortical ribbon, it is possible to obtain a global 3D gyrification index based on the ratio between the surface area of the cortical surface and that of the convex envelope of the cortex. In addition, the 2D spatial normalization of the cortex surface (Fischl, 2012) can be used to obtain a similar local gyrification index, calculated at each point, but which is not very easy to interpret. The global gyrification index can also be enriched by quantifying the amount of energy required by the folding of a cortex for different frequency bands (Germanaud et al., 2014), a bit like an EEG signal decomposition into alpha, beta and theta waves. This approach is of particular interest for the study of development, as there is a correspondence between these bands and primary, secondary and tertiary folding (Dubois et al., 2019).

To overcome the difficulties associated with spatial normalization when studying cortical morphology, it is sometimes preferable to return to the approach of ROI. One of the software distributed in the community, brainVISA, allows the segmentation of more than a hundred sulci (Perrot, Riviere, & Mangin, 2011). These virtual objects correspond to a negative 3D print of the folds of the cortex. They are automatically identified using an artificial intelligence approach, trained on a database where they have been identified by human experts. Then, for each of the sulci, it is possible to quantify simple morphometric parameters such as the length, depth or distance between the two walls, a marker of local atrophy (Mangin et al., 2010). We can then follow the evolution of these parameters from the onset of the sulcus to adulthood. It is also possible to test whether pathologies of developmental origin have an impact on these parameters. To go further, more sophisticated sulcus shape characterization, such as qualitative features (e.g. sulcus interruptions, absence/presence of a fold) are of interest, but they still require visual observation to be reliable. Unsupervised machine learning approaches under development are in the process of developing a dictionary of the most frequent local folding patterns in the general population. This strategy will allow better modeling of the normal variability of cortical folding, which should lead to automated techniques for anomaly detection. Each type of abnormality is likely to indicate a specific developmental event impacting brain architecture. These abnormalities could assist in patient stratification.

One of the major difficulties of the ROI approach focusing on cortical sulci lies in the weaknesses of their definition in the anatomical literature. Large sulci are often interrupted, each sulcus piece being susceptible to connecting with the others in various ways. This recombination process often leads to ambiguous configurations for the usual anatomical nomenclature, which creates difficulties for the morphometric study of sulci. These difficulties have led to propose an alternative nomenclature of cortical folding defined at a lower scale level (Mangin et al., 2019). This nomenclature is based on unbreakable entities called ‘sulcal roots’ supposedly present in each brain. These sulcal roots can be separated by transverse gyri historically called "passage folds" often buried in the depths of the grooves (see Fig Box2). At present, the best way to project this nomenclature into a brain is to rely on local maxima of depth of the cortical surface called sulcal pits (Lohmann, von Cramon, & Colchester, 2008). Several studies have shown good reproducibility of the pits map across individuals and ages. One of the current research programs is to understand the genesis of these pits in antenatal images and their possible links with early brain architecture. Several studies have shown differences in the organization of the pits map in the context of abnormal development.

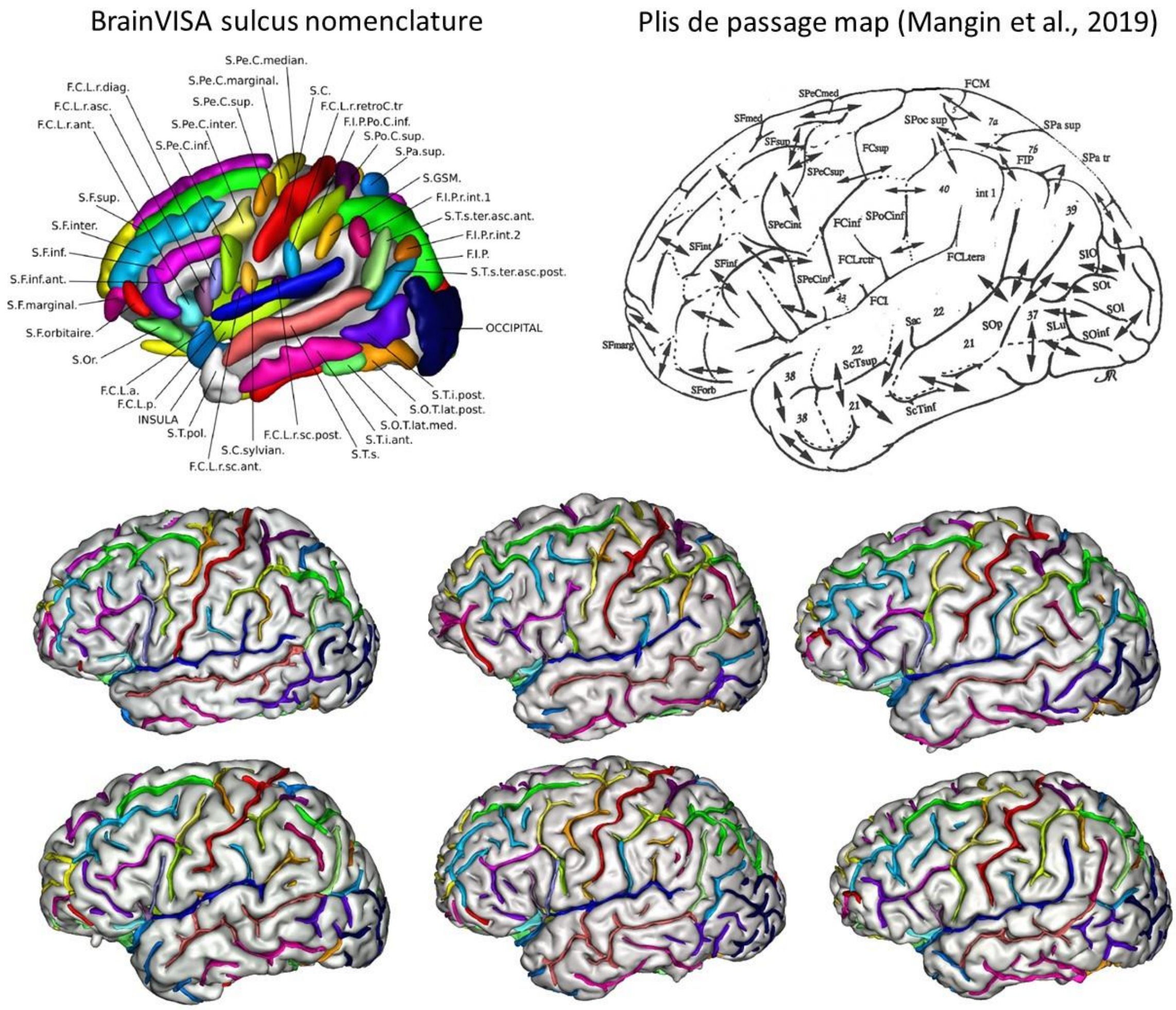

***Figure 5: Top: Sulcus nomenclature and map of the standard sulcus interruptions. Bottom: The sulcus nomenclature projected on six different left hemispheres.***

### 6.3. BOX 3: *Measuring the white matter microstructure and connectivity with diffusion MRI*

Conventional anatomical MRI can easily distinguish grey matter from white matter (WM), but provides little information about their microscopic structure. Diffusion MRI (dMRI), on the other hand, exploits the random motions of water in tissue to provide microstructure features that were previously only accessible by histology (Johansen-Berg & Behrens, 2013). Diffusion MRI reveals regularities in the microscopic geometry of cells through its impact on the diffusion of water molecules (the diffusion process is disturbed by the cell membranes). These perturbations lead to MR signal heterogeneities, for example in the extent of the diffusion process in each direction of space. Typically, in white matter, the amplitude of this process is greater in the direction of the fibers, whereas it is hindered and restricted in transverse directions of the fibers (Figure 6a).

This local anisotropy of the water diffusion process in WM has led to a flagship application of diffusion MRI: the mapping of structural connectivity. This application has major impact since dissection only allows the observation of the largest WM pathways, whereas diffusion MRI can reveal, for the first time, the large pathways but also short connections such as U-shaped fibers in vivo (Guevara et al., 2017). Over the last twenty years, a wide variety of models have been proposed to decode the diffusion signal in each MRI voxel. The aim is to estimate the direction of the WM fibers locally in order to be able to reconstruct the trajectory of the fascicles step by step (Figure 6b). The major difficulty lies in the complexity of the geometry of the fiber crossings, which constitute most of the white matter. The most recent approaches, based on acquisition sequences that probe the diffusion of water in a large number of spatial directions, now make it possible to decipher a large part of these crossings, even though ambiguous situations still remain. It is nevertheless possible today to map more than a hundred bundles to study their integrity or their maturation. Dedicated measurements are derived from the diffusion of water for this purpose. They are averaged either throughout the bundle or along a central axis (Figure 6c). The Fractional Anisotropy (FA), which quantifies the amplitude of diffusion in the direction of the fibres versus in transverse directions, is generally maximal when a bundle is in good condition: the membranes and myelin sheaths that disrupt diffusion constitute barriers that are difficult to cross. On the other hand, when the bundle becomes disorganized, for example due to degeneration, FA decreases. The same type of considerations can be used to interpret diffusion variations induced by brain development and fibre myelination.

A range of recent microstructure imaging techniques are proposing to tackle various histological features (Alexander, Dyrby, Nilsson, & Zhang, 2019): neurite (axon or dendrite) density, axon diameter distributions, cell shape, myelin density, myelin water fraction, etc. These techniques often combine diffusion imaging with quantitative relaxometry (the estimation of T1 and T2 relaxation times at the core of magnetic resonance). Most of these techniques fit a model relating the microscopic tissue features to MR signals in each voxel to produce microstructure maps. For instance, axon diameter stems from a simple model consisting of parallel impermeable cylinders, which is fitted in each voxel to a set of varying diffusion weighting in order to recover estimates of cylinder size and packing. Current models have several limitations, mainly related to the few model parameters that can be estimated with practical acquisition protocols. While these advanced technologies are still in progress, they have great appeal for studying brain development and maturation.

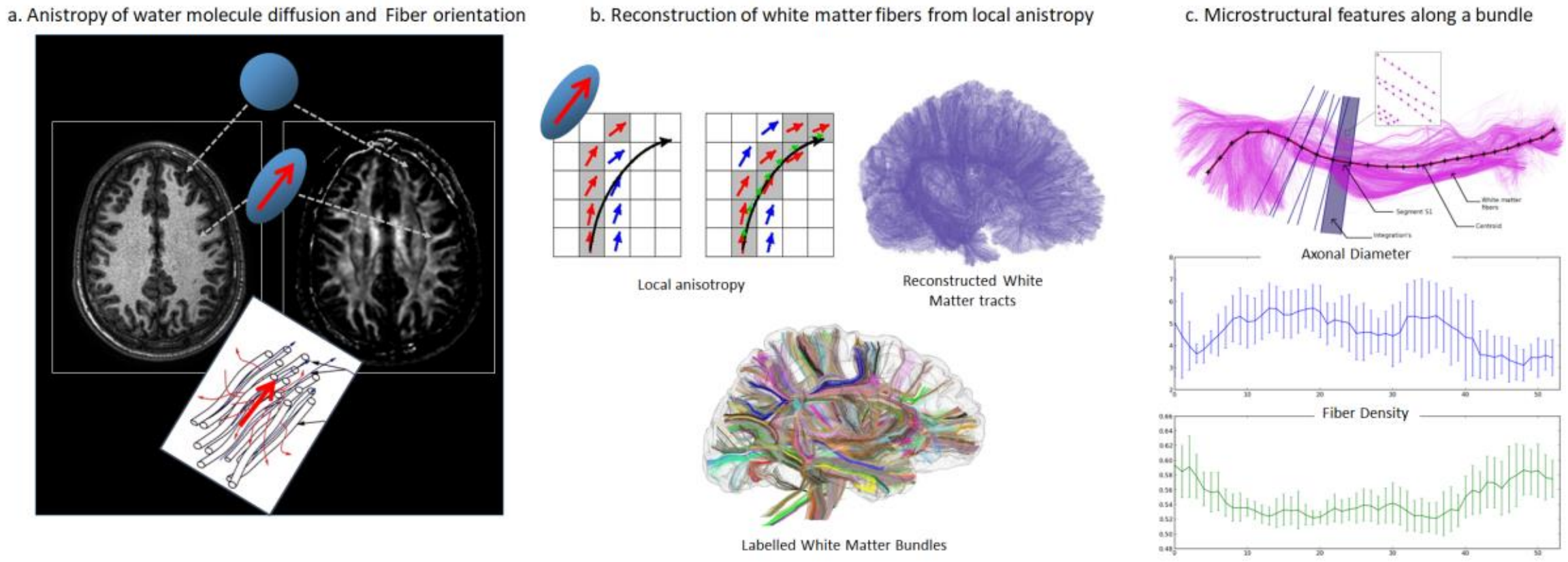

***Figure Box3: white matter microstructure and connectivity with diffusion MRI.*** *a. Water molecule diffusion is isotropic (sphere) in grey matter and anisotropic (ellipsoid) in grey matter. The main direction of water molecule diffusion (red arrow) indicate the main direction of the white matter fibers. b. Whole white matter tracts can be mathematically reconstructed step-by-step using local tract direction derived from diffusion anisotropy. Main white matter bundles can be identified from all reconstructed tracts. c. Microstructural features (e.g. axonal diameter, fiber density) can be estimated along each white matter bundle.*

## Acknowledgment

Arnaud Cachia was supported by the “Agence National pour la Recherche", "Institut Universitaire de France" and the "Fondation pour la Recherche Médicale". Jean-François Mangin was supported by the European Union’s HBP-SGA2 (grant agreement n°785907), FRM (DIC20161236445), ANR (IFOPASUBA), and the Blaise Pascal Chair of Region Ile de France and Université Paris-Saclay to W. Hopkins. Jessica Dubois was supported by the “Fondation de France”, the Fyssen Foundation, the Médisite Foundation and the IdEx Université de Paris (ANR-18-IDEX-0001).

## References

Adibpour, P. , Lebenberg, J., Kabdebon, C., Dehaene-Lambertz, G., & Dubois, J. (2020). Anatomo-functional correlates of auditory development in infancy. *Developmental Cognitive Neuroscience*.

Adibpour, P., Dehaene-Lambertz, G., & Dubois, J. (2015). Relating the structural and functional maturation of visual and auditory white matter pathways with diffusion imaging and event-related potentials in infants. *Proceedings of ISMRM meeting*, 645.

Adibpour, P., Dubois, J., & Dehaene-Lambertz, G. (2018). Right but not left hemispheric discrimination of faces in infancy. *Nat Hum Behav*, *2*(1), 67-79. doi: 10.1038/s41562-017-0249-4

Adibpour, P., Dubois, J., Moutard, M. L., & Dehaene-Lambertz, G. (2018). Early asymmetric inter-hemispheric transfer in the auditory network: insights from infants with corpus callosum agenesis. *Brain Struct Funct, 223*(6), 2893-2905. doi: 10.1007/s00429-018-1667-4

Alexander, Daniel C, Dyrby, Tim B, Nilsson, Markus, & Zhang, Hui. (2019). Imaging brain microstructure with diffusion MRI: practicality and applications. *NMR in Biomedicine, 32*(4), e3841.

Amiez, C., Wilson, C. R. E., & Procyk, E. (2018). Variations of cingulate sulcal organization and link with cognitive performance. *Sci Rep, 8*(1), 13988. doi: 10.1038/s41598-018-32088-9

Anderson, V., Spencer-Smith, M., & Wood, A. (2011). Do children really recover better? Neurobehavioural plasticity after early brain insult. *Brain, 134*(Pt 8), 2197-2221. doi: 10.1093/brain/awr103

Andescavage, N. N., du Plessis, A., McCarter, R., Serag, A., Evangelou, I., Vezina, G., Robertson, R., & Limperopoulos, C. (2017). Complex Trajectories of Brain Development in the Healthy Human Fetus. *Cereb Cortex, 27*(11), 5274-5283. doi: 10.1093/cercor/bhw306

Angleys, H., Germanaud, D., Leroy, F., Hertz-Pannier, L., Mangin, J.F., Lazeyras, F., Hüppi, P.S., Lefèvre, J., & Dubois, J. (2014). Successive waves of cortical folding in the developing brain using spectral analysis of gyrification. *Proceedings of OHBM*.

Ashburner, J., & Friston, K. J. (2000). Voxel-based morphometry--the methods. *Neuroimage, 11*(6 Pt 1), 805-821.

Bajic, D., Wang, C., Kumlien, E., Mattsson, P., Lundberg, S., Eeg-Olofsson, O., & Raininko, R. (2008). Incomplete inversion of the hippocampus--a common developmental anomaly. *Eur Radiol, 18*(1), 138-142. doi: 10.1007/s00330-007-0735-6

Ball, G., Aljabar, P., Zebari, S., Tusor, N., Arichi, T., Merchant, N., Robinson, E. C., Ogundipe, E., Rueckert, D., Edwards, A. D., & Counsell, S. J. (2014). Rich-club organization of the newborn human brain. *Proc Natl Acad Sci U S A, 111*(20), 7456-7461. doi: 1324118111 [pii] 10.1073/pnas.1324118111

Ball, G., Pazderova, L., Chew, A., Tusor, N., Merchant, N., Arichi, T., Allsop, J. M., Cowan, F. M., Edwards, A. D., & Counsell, S. J. (2015). Thalamocortical Connectivity Predicts Cognition in Children Born Preterm. *Cereb Cortex, 25*, 4310-4318. doi: bhu331 [pii] 10.1093/cercor/bhu331

Ball, G., Srinivasan, L., Aljabar, P., Counsell, S. J., Durighel, G., Hajnal, J. V., Rutherford, M. A., & Edwards, A. D. (2013). Development of cortical microstructure in the preterm human brain. *Proc Natl Acad Sci U S A, 110*(23), 9541-9546. doi: 1301652110 [pii] 10.1073/pnas.1301652110

Barkovich, A. J., Kjos, B. O., Jackson, D. E., Jr., & Norman, D. (1988). Normal maturation of the neonatal and infant brain: MR imaging at 1.5 T. *Radiology, 166*(1 Pt 1), 173-180.

Batalle, D., O'Muircheartaigh, J., Makropoulos, A., Kelly, C. J., Dimitrova, R., Hughes, E. J., Hajnal, J. V., Zhang, H., Alexander, D. C., David Edwards, A., & Counsell, S. J. (2018). Different patterns of cortical maturation before and after 38 weeks gestational age demonstrated by diffusion MRI in vivo. *Neuroimage*. doi: S1053-8119(18)30460-9 [pii] 10.1016/j.neuroimage.2018.05.046

Baumann, N., & Pham-Dinh, D. (2001). Biology of oligodendrocyte and myelin in the mammalian central nervous system. *Physiol Rev, 81*(2), 871-927.

Betzel, Richard F, & Bassett, Danielle S. (2017). Multi-scale brain networks. *Neuroimage, 160*, 73-83.

Borrell, V. (2018). How Cells Fold the Cerebral Cortex. *J Neurosci, 38*(4), 776-783. doi: 10.1523/JNEUROSCI.1106-17.2017

Borst, G, Cachia, A, Tissier, C, Ahr, E., Simon, G, & Houdé, O (2016). Early cerebral constraint on reading skills of 10-years-old children. *Mind, Brain and Education, 10*(1), 47-54.

Borst, G., Cachia, A., Vidal, J., Simon, G., Fischer, C., Pineau, A., Poirel, N., Mangin, J. F., & Houde, O. (2014). Folding of the anterior cingulate cortex partially explains inhibitory control during childhood: a longitudinal study. *Dev Cogn Neurosci, 9*, 126-135. doi: 10.1016/j.dcn.2014.02.006

Bozek, J., Makropoulos, A., Schuh, A., Fitzgibbon, S., Wright, R., Glasser, M. F., Coalson, T. S., O'Muircheartaigh, J., Hutter, J., Price, A. N., Cordero-Grande, L., Teixeira, Rpag, Hughes, E., Tusor, N., Baruteau, K. P., Rutherford, M. A., Edwards, A. D., Hajnal, J. V., Smith, S. M., Rueckert, D., Jenkinson, M., & Robinson, E. C. (2018). Construction of a neonatal cortical surface atlas using Multimodal Surface Matching in the Developing Human Connectome Project. *Neuroimage, 179*, 11-29. doi: S1053-8119(18)30525-1 [pii] 10.1016/j.neuroimage.2018.06.018

Brody, B. A., Kinney, H. C., Kloman, A. S., & Gilles, F. H. (1987). Sequence of central nervous system myelination in human infancy. I. An autopsy study of myelination. *J Neuropathol Exp Neurol, 46*(3), 283-301.

Brown, C. J., Miller, S. P., Booth, B. G., Andrews, S., Chau, V., Poskitt, K. J., & Hamarneh, G. (2014). Structural network analysis of brain development in young preterm neonates. *Neuroimage, 101*, 667-680. doi: S1053-8119(14)00613-2 [pii] 10.1016/j.neuroimage.2014.07.030

Bui, T., Daire, J. L., Chalard, F., Zaccaria, I., Alberti, C., Elmaleh, M., Garel, C., Luton, D., Blanc, N., & Sebag, G. (2006). Microstructural development of human brain assessed in utero by diffusion tensor imaging. *Pediatr Radiol, 36*(11), 1133-1140. doi: 10.1007/s00247-006-0266-3

Bultmann, E., Spineli, L. M., Hartmann, H., & Lanfermann, H. (2018). Measuring in vivo cerebral maturation using age-related T2 relaxation times at 3T. *Brain Dev, 40*, 85-93. doi: S0387-7604(17)30208-5 [pii] 10.1016/j.braindev.2017.07.011

Cachia, A., Borst, G., Tissier, C., Fisher, C., Plaze, M., Gay, O., Riviere, D., Gogtay, N., Giedd, J., Mangin, J. F., Houde, O., & Raznahan, A. (2016). Longitudinal stability of the folding pattern of the anterior cingulate cortex during development. *Dev Cogn Neurosci, 19*, 122-127. doi: 10.1016/j.dcn.2016.02.011

Cachia, A., Borst, G., Vidal, J., Fischer, C., Pineau, A., Mangin, J. F., & Houde, O. (2014). The shape of the ACC contributes to cognitive control efficiency in preschoolers. *J Cogn Neurosci, 26*(1), 96-106. doi: 10.1162/jocn_a_00459

Cachia, A., Del Maschio, N., Borst, G., Della Rosa, P. A., Pallier, C., Costa, A., Houde, O., & Abutalebi, J. (2017). Anterior cingulate cortex sulcation and its differential effects on conflict monitoring in bilinguals and monolinguals. *Brain Lang, 175*, 57-63. doi: 10.1016/j.bandl.2017.09.005

Cachia, A., Roell, M., Mangin, J. F., Sun, Z. Y., Jobert, A., Braga, L., Houde, O., Dehaene, S., & Borst, G. (2017). How interindividual differences in brain anatomy shape reading accuracy. *Brain Struct Funct*. doi: 10.1007/s00429-017-1516-x 10.1007/s00429-017-1516-x [pii]

Cachia, A., Roell, M., Mangin, J. F., Sun, Z. Y., Jobert, A., Braga, L., Houde, O., Dehaene, S., & Borst, G. (2018). How interindividual differences in brain anatomy shape reading accuracy. *Brain Struct Funct, 223*(2), 701-712. doi: 10.1007/s00429-017-1516-x

Cao, M., Huang, H., & He, Y. (2017). Developmental Connectomics from Infancy through Early Childhood. *Trends Neurosci, 40*, 494-506. doi: S0166-2236(17)30115-7 [pii] 10.1016/j.tins.2017.06.003

Chevalier, N., Kurth, S., Doucette, M. R., Wiseheart, M., Deoni, S. C., Dean, D. C., 3rd, O'Muircheartaigh, J., Blackwell, K. A., Munakata, Y., & LeBourgeois, M. K. (2015). Myelination Is Associated with Processing Speed in Early Childhood:

Preliminary Insights. *PLoS One, 10*(10), e0139897. doi: 10.1371/journal.pone.0139897 PONE-D-15-02602 [pii]

Chi, J. G., Dooling, E. C., & Gilles, F. H. (1977). Gyral development of the human brain. *Ann Neurol, 1*(1), 86-93.

Cohen, Alexander L, Fair, Damien A, Dosenbach, Nico UF, Miezin, Francis M, Dierker, Donna, Van Essen, David C, Schlaggar, Bradley L, & Petersen, Steven E. (2008). Defining functional areas in individual human brains using resting functional connectivity MRI. *Neuroimage, 41*(1), 45-57.

Croteau-Chonka, E. C., Dean, D. C., 3rd, Remer, J., Dirks, H., O'Muircheartaigh, J., & Deoni, S. C. (2016). Examining the relationships between cortical maturation and white matter myelination throughout early childhood. *Neuroimage, 125*, 413-421. doi: S1053-8119(15)00949-0 [pii] 10.1016/j.neuroimage.2015.10.038

Cury, C., Toro, R., Cohen, F., Fischer, C., Mhaya, A., Samper-Gonzalez, J., Hasboun, D., Mangin, J. F., Banaschewski, T., Bokde, A. L., Bromberg, U., Buechel, C., Cattrell, A., Conrod, P., Flor, H., Gallinat, J., Garavan, H., Gowland, P., Heinz, A., Ittermann, B., Lemaitre, H., Martinot, J. L., Nees, F., Paillere Martinot, M. L., Orfanos, D. P., Paus, T., Poustka, L., Smolka, M. N., Walter, H., Whelan, R., Frouin, V., Schumann, G., Glaunes, J. A., Colliot, O., & Consortium, Imagen. (2015). Incomplete Hippocampal Inversion: A Comprehensive MRI Study of Over 2000 Subjects. *Front Neuroanat, 9*, 160. doi: 10.3389/fnana.2015.00160

De Guio, F., Mangin, J. F., Riviere, D., Perrot, M., Molteno, C. D., Jacobson, S. W., Meintjes, E. M., & Jacobson, J. L. (2014). A study of cortical morphology in children with fetal alcohol spectrum disorders. *Hum Brain Mapp, 35*(5), 2285-2296. doi: 10.1002/hbm.22327

Dean, D. C., 3rd, O'Muircheartaigh, J., Dirks, H., Travers, B. G., Adluru, N., Alexander, A. L., & Deoni, S. C. (2016). Mapping an index of the myelin g-ratio in infants using magnetic resonance imaging. *Neuroimage, 132*, 225-237. doi: S1053-8119(16)00149-X [pii] 10.1016/j.neuroimage.2016.02.040

Dean, D. C., 3rd, Planalp, E. M., Wooten, W., Adluru, N., Kecskemeti, S. R., Frye, C., Schmidt, C. K., Schmidt, N. L., Styner, M. A., Goldsmith, H. H., Davidson, R. J., & Alexander, A. L. (2017). Mapping White Matter Microstructure in the One Month Human Brain. *Sci Rep, 7*(1), 9759. doi: 10.1038/s41598-017-09915-6 10.1038/s41598-017-09915-6 [pii]

Dehaene-Lambertz, G., & Spelke, E. S. (2015). The Infancy of the Human Brain. *Neuron, 88*(1), 93-109. doi: 10.1016/j.neuron.2015.09.026

Deipolyi, A. R., Mukherjee, P., Gill, K., Henry, R. G., Partridge, S. C., Veeraraghavan, S., Jin, H., Lu, Y., Miller, S. P., Ferriero, D. M., Vigneron, D. B., & Barkovich, A. J. (2005). Comparing microstructural and macrostructural development of the cerebral cortex in premature newborns: diffusion tensor imaging versus cortical gyration. *Neuroimage, 27*(3), 579-586. doi: S1053-8119(05)00275-2 [pii] 10.1016/j.neuroimage.2005.04.027

Del Maschio, N., Sulpizio, S., Fedeli, D., Ramanujan, K., Ding, G., Weekes, B. S., Cachia, A., & Abutalebi, J. (2018). ACC Sulcal Patterns and Their Modulation on Cognitive Control Efficiency Across Lifespan: A Neuroanatomical Study on Bilinguals and Monolinguals. *Cereb Cortex*. doi: 10.1093/cercor/bhy175

Deoni, S. C., Dean, D. C., 3rd, O'Muircheartaigh, J., Dirks, H., & Jerskey, B. A. (2012). Investigating white matter development in infancy and early childhood using myelin water faction and relaxation time mapping. *Neuroimage, 63*(3), 1038-1053. doi: S1053-8119(12)00766-5 [pii] 10.1016/j.neuroimage.2012.07.037

Deoni, S. C., Dean, D. C., 3rd, Remer, J., Dirks, H., & O'Muircheartaigh, J. (2015). Cortical maturation and myelination in healthy toddlers and young children. *Neuroimage, 115*, 147-161. doi: S1053-8119(15)00358-4 [pii] 10.1016/j.neuroimage.2015.04.058

Deoni, S. C., Mercure, E., Blasi, A., Gasston, D., Thomson, A., Johnson, M., Williams, S. C., & Murphy, D. G. (2011). Mapping infant brain myelination with magnetic resonance imaging. *J Neurosci, 31*(2), 784-791. doi: 31/2/784 [pii] 10.1523/JNEUROSCI.2106-10.2011

Deoni, S. C., O'Muircheartaigh, J., Elison, J. T., Walker, L., Doernberg, E., Waskiewicz, N., Dirks, H., Piryatinsky, I., Dean, D. C., 3rd, & Jumbe, N. L. (2016). White matter maturation profiles through early childhood predict general cognitive ability. *Brain Struct Funct, 221*, 1189-1203. doi: 10.1007/s00429-014-0947-x

Dockstader, C., Gaetz, W., Rockel, C., & Mabbott, D. J. (2012). White matter maturation in visual and motor areas predicts the latency of visual activation in children. *Hum Brain Mapp, 33*(1), 179-191. doi: 10.1002/hbm.21203

Dubois, J., Adibpour, P., Poupon, C., Hertz-Pannier, L., & Dehaene-Lambertz, G. (2016). MRI and M/EEG studies of the White Matter Development in Human Fetuses and Infants: Review and Opinion. *Brain Plast, 2*(1), 49-69. doi: 10.3233/BPL-160031

Dubois, J., Benders, M., Borradori-Tolsa, C., Cachia, A., Lazeyras, F., Ha-Vinh Leuchter, R., Sizonenko, S. V., Warfield, S. K., Mangin, J. F., & Huppi, P. S. (2008). Primary cortical folding in the human newborn: an early marker of later functional development. *Brain, 131*(Pt 8), 2028-2041.

Dubois, J., Benders, M., Cachia, A., Lazeyras, F., Leuchter, R. H. V., Sizonenko, S. V., Borradori-Tolsa, C., Mangin, J. F., & Huppi, P. S. (2008). Mapping the early cortical folding process in the preterm newborn brain. *Cerebral Cortex, 18*(6), 1444-1454. doi: 10.1093/cercor/bhm180

Dubois, J., & Dehaene-Lambertz, G. (2015). Fetal and postnatal development of the cortex: insights from MRI and genetics. In Arthur W. Toga (Ed.), *Brain Mapping: An Encyclopedic Reference* (Vol. 2, pp. 11-19): Academic Press: Elsevier.

Dubois, J., Dehaene-Lambertz, G., Kulikova, S., Poupon, C., Huppi, P. S., & Hertz-Pannier, L. (2014). The early development of brain white matter: a review of imaging studies in fetuses, newborns and infants. *Neuroscience, 276*, 48-71. doi: 10.1016/j.neuroscience.2013.12.044

Dubois, J., Dehaene-Lambertz, G., Perrin, M., Mangin, J. F., Cointepas, Y., Duchesnay, E., Le Bihan, D., & Hertz-Pannier, L. (2008). Asynchrony of the early maturation of white matter bundles in healthy infants: quantitative landmarks revealed noninvasively by diffusion tensor imaging. *Hum Brain Mapp, 29*(1), 14-27. doi: 10.1002/hbm.20363

Dubois, J., Dehaene-Lambertz, G., Soares, C., Cointepas, Y., Le Bihan, D., & Hertz-Pannier, L. (2008). Microstructural correlates of infant functional development: example of the visual pathways. *J Neurosci, 28*(8), 1943-1948. doi: 10.1523/JNEUROSCI.5145-07.2008

Dubois, J., Hertz-Pannier, L., Dehaene-Lambertz, G., Cointepas, Y., & Le Bihan, D. (2006). Assessment of the early organization and maturation of infants' cerebral white matter fiber bundles: a feasibility study using quantitative diffusion tensor imaging and tractography. *Neuroimage, 30*(4), 1121-1132.

Dubois, J., Kostovic, I., & Judas, M. . (2015). Development of structural and functional connectivity. In Arthur W. Toga (Ed.), *Brain Mapping: An Encyclopedic Reference* (Vol. 2, pp. 423-437): Academic Press: Elsevier.

Dubois, J., Lefevre, J., Angleys, H., Leroy, F., Fischer, C., Lebenberg, J., Dehaene-Lambertz, G., Borradori-Tolsa, C., Lazeyras, F., Hertz-Pannier, L., Mangin, J. F., Huppi, P. S., & Germanaud, D. (2019). The dynamics of cortical folding waves and prematurity-related deviations revealed by spatial and spectral analysis of gyrification. *Neuroimage, 185*, 934-946. doi: 10.1016/j.neuroimage.2018.03.005

Dubois, J., Poupon, C., Thirion, B., Simonnet, H., Kulikova, S., Leroy, F., Hertz-Pannier, L., & Dehaene-Lambertz, G. (2016). Exploring the Early Organization and Maturation of Linguistic Pathways in the Human Infant Brain. *Cereb Cortex, 26*(5), 2283-2298. doi: 10.1093/cercor/bhv082

Dudink, J., Buijs, J., Govaert, P., van Zwol, A. L., Conneman, N., van Goudoever, J. B., & Lequin, M. (2010). Diffusion tensor imaging of the cortical plate and subplate in very-low-birth-weight infants. *Pediatr Radiol, 40*(8), 1397-1404. doi: 10.1007/s00247-010-1638-2

Dudink, J., Lequin, M., van Pul, C., Buijs, J., Conneman, N., van Goudoever, J., & Govaert, P. (2007). Fractional anisotropy in white matter tracts of very-low-birth-weight infants. *Pediatr Radiol, 37*(12), 1216-1223. doi: 10.1007/s00247-007-0626-7

Eaton-Rosen, Z., Melbourne, A., Orasanu, E., Cardoso, M. J., Modat, M., Bainbridge, A., Kendall, G. S., Robertson, N. J., Marlow, N., & Ourselin, S. (2015). Longitudinal measurement of the developing grey matter in preterm subjects using multi-modal MRI. *Neuroimage, 111*, 580-589. doi: S1053-8119(15)00102-0 [pii] 10.1016/j.neuroimage.2015.02.010

Eaton-Rosen, Z., Scherrer, B., Melbourne, A., Ourselin, S., Neil, J. J., & Warfield, S. K. (2017). Investigating the maturation of microstructure and radial orientation in the preterm human cortex with diffusion MRI. *Neuroimage, 162*, 65-72. doi: S1053-8119(17)30659-6 [pii] 10.1016/j.neuroimage.2017.08.013

Eickhoff, S., Walters, N. B., Schleicher, A., Kril, J., Egan, G. F., Zilles, K., Watson, J. D., & Amunts, K. (2005). High-resolution MRI reflects myeloarchitecture and cytoarchitecture of human cerebral cortex. *Hum Brain Mapp, 24*(3), 206-215. doi: 10.1002/hbm.20082

Engelbrecht, V., Rassek, M., Preiss, S., Wald, C., & Modder, U. (1998). Age-dependent changes in magnetization transfer contrast of white matter in the pediatric brain. *AJNR Am J Neuroradiol, 19*(10), 1923-1929.

Fair, Damien A, Cohen, Alexander L, Power, Jonathan D, Dosenbach, Nico UF, Church, Jessica A, Miezin, Francis M, Schlaggar, Bradley L, & Petersen, Steven E. (2009). Functional brain networks develop from a "local to distributed" organization. *PLoS computational biology, 5*(5), e1000381.

Fair, Damien A, Dosenbach, Nico UF, Church, Jessica A, Cohen, Alexander L, Brahmbhatt, Shefali, Miezin, Francis M, Barch, Deanna M, Raichle, Marcus E, Petersen, Steven E, & Schlaggar, Bradley L. (2007). Development of distinct control networks through segregation and integration. *Proceedings of the National Academy of Sciences, 104*(33), 13507-13512.

Feess-Higgins, A., & Larroche, JC. (1987). *Development of the human foetal brain. An anatomical atlas.* (Masson ed.): INSERM CNRS.

Ferradal, S. L., Gagoski, B., Jaimes, C., Yi, F., Carruthers, C., Vu, C., Litt, J. S., Larsen, R., Sutton, B., Grant, P. E., & Zollei, L. (2018). System-specific patterns of thalamocortical connectivity in early brain development as revealed by structural and functional MRI. *Cereb Cortex*. doi: 4841689 [pii] 10.1093/cercor/bhy028

Fields, R. D. (2008). White matter in learning, cognition and psychiatric disorders. *Trends Neurosci, 31*(7), 361-370.

Fischl, B. (2012). FreeSurfer. *Neuroimage, 62*(2), 774-781. doi: 10.1016/j.neuroimage.2012.01.021

Flechsig, P. (1920). Anatomie des Menschlichen Gehirn und Rückenmarks, auf myelogenetischer grundlage. *G. Thieme*.

Fornito, A., Yucel, M., Wood, S., Stuart, G. W., Buchanan, J. A., Proffitt, T., Anderson, V., Velakoulis, D., & Pantelis, C. (2004). Individual differences in anterior cingulate/paracingulate morphology are related to executive functions in healthy males. *Cereb Cortex, 14*(4), 424-431.

Foubet, O., Trejo, M., & Toro, R. (2018). Mechanical morphogenesis and the development of neocortical organisation. *Cortex*. doi: 10.1016/j.cortex.2018.03.005

Friedrichs-Maeder, C. L., Griffa, A., Schneider, J., Huppi, P. S., Truttmann, A., & Hagmann, P. (2017). Exploring the role of white matter connectivity in cortex

maturation. *PLoS One, 12*(5), e0177466. doi: 10.1371/journal.pone.0177466 PONE-D-16-40936 [pii]

Geng, X., Gouttard, S., Sharma, A., Gu, H., Styner, M., Lin, W., Gerig, G., & Gilmore, J. H. (2012). Quantitative tract-based white matter development from birth to age 2years. *Neuroimage, 61*(3), 542-557. doi: S1053-8119(12)00338-2 [pii] 10.1016/j.neuroimage.2012.03.057

Germanaud, D., Lefevre, J., Fischer, C., Bintner, M., Curie, A., des Portes, V., Eliez, S., Elmaleh-Berges, M., Lamblin, D., Passemard, S., Operto, G., Schaer, M., Verloes, A., Toro, R., Mangin, J. F., & Hertz-Pannier, L. (2014). Simplified gyral pattern in severe developmental microcephalies? New insights from allometric modeling for spatial and spectral analysis of gyrification. *Neuroimage, 102 Pt 2*, 317-331. doi: 10.1016/j.neuroimage.2014.07.057

Giedd, J. N., & Rapoport, J. L. (2010). Structural MRI of pediatric brain development: what have we learned and where are we going? *Neuron, 67*(5), 728-734. doi: 10.1016/j.neuron.2010.08.040

Gilles, F., Shankle, W., & Dooling, E. (1983). Myelinated tracts: growth patterns. *Gilles F., Leviton A. and Dooling E. eds., John Wright PSG, Boston.*

Gilmore, J. H., Lin, W., Prastawa, M. W., Looney, C. B., Vetsa, Y. S., Knickmeyer, R. C., Evans, D. D., Smith, J. K., Hamer, R. M., Lieberman, J. A., & Gerig, G. (2007). Regional gray matter growth, sexual dimorphism, and cerebral asymmetry in the neonatal brain. *J Neurosci, 27*(6), 1255-1260. doi: 27/6/1255 [pii] 10.1523/JNEUROSCI.3339-06.2007

Gilmore, J. H., Shi, F., Woolson, S. L., Knickmeyer, R. C., Short, S. J., Lin, W., Zhu, H., Hamer, R. M., Styner, M., & Shen, D. (2012). Longitudinal Development of Cortical and Subcortical Gray Matter from Birth to 2 Years. *Cereb Cortex*. doi: bhr327 [pii] 10.1093/cercor/bhr327

Guevara, Miguel, Román, Claudio, Houenou, Josselin, Duclap, Delphine, Poupon, Cyril, Mangin, Jean François, & Guevara, Pamela. (2017). Reproducibility of superficial white matter tracts using diffusion-weighted imaging tractography. *Neuroimage, 147*, 703-725.

Habas, P. A., Scott, J. A., Roosta, A., Rajagopalan, V., Kim, K., Rousseau, F., Barkovich, A. J., Glenn, O. A., & Studholme, C. (2012). Early folding patterns and asymmetries of the normal human brain detected from in utero MRI. *Cereb Cortex, 22*(1), 13-25. doi: bhr053 [pii] 10.1093/cercor/bhr053

Harding, G. F., Grose, J., Wilton, A., & Bissenden, J. G. (1989). The pattern reversal VEP in short-gestation infants. *Electroencephalogr Clin Neurophysiol, 74*(1), 76-80.

Haselgrove, J., Moore, J., Wang, Z., Traipe, E., & Bilaniuk, L. (2000). A method for fast multislice T1 measurement: feasibility studies on phantoms, young children, and children with Canavan's disease. *J Magn Reson Imaging, 11*(4), 360-367. doi: 10.1002/(SICI)1522-2586(200004)11:4<360::AID-JMRI3>3.0.CO;2-G [pii]

Hilgetag, C. C., & Barbas, H. (2006). Role of mechanical factors in the morphology of the primate cerebral cortex. *PLoS Comput Biol, 2*(3), e22. doi: 10.1371/journal.pcbi.0020022

Hilgetag, C. C., & Barbas, H. (2009). Sculpting the brain. *Sci Am, 300*(2), 66-71.

Hill, J., Dierker, D., Neil, J., Inder, T., Knutsen, A., Harwell, J., Coalson, T., & Van Essen, D. (2010). A surface-based analysis of hemispheric asymmetries and folding of cerebral cortex in term-born human infants. *J Neurosci, 30*(6), 2268-2276. doi: 30/6/2268 [pii]

10.1523/JNEUROSCI.4682-09.2010

Horowitz, A., Barazany, D., Tavor, I., Bernstein, M., Yovel, G., & Assaf, Y. (2015). In vivo correlation between axon diameter and conduction velocity in the human brain. *Brain Struct Funct, 220*(3), 1777-1788. doi: 10.1007/s00429-014-0871-0

Huang, H., Jeon, T., Sedmak, G., Pletikos, M., Vasung, L., Xu, X., Yarowsky, P., Richards, L. J., Kostovic, I., Sestan, N., & Mori, S. (2013). Coupling Diffusion Imaging with Histological and Gene Expression Analysis to Examine the Dynamics of Cortical Areas across the Fetal Period of Human Brain Development. *Cereb Cortex, 23*, 2620-2631. doi: bhs241 [pii] 10.1093/cercor/bhs241

Huang, H., Xue, R., Zhang, J., Ren, T., Richards, L. J., Yarowsky, P., Miller, M. I., & Mori, S. (2009). Anatomical characterization of human fetal brain development with diffusion tensor magnetic resonance imaging. *J Neurosci, 29*(13), 4263-4273. doi: 29/13/4263 [pii] 10.1523/JNEUROSCI.2769-08.2009

Huang, H., Zhang, J., Wakana, S., Zhang, W., Ren, T., Richards, L. J., Yarowsky, P., Donohue, P., Graham, E., van Zijl, P. C., & Mori, S. (2006). White and gray matter development in human fetal, newborn and pediatric brains. *Neuroimage, 33*(1), 27-38. doi: S1053-8119(06)00670-7 [pii] 10.1016/j.neuroimage.2006.06.009

Huppi, P. S., & Dubois, J. (2006). Diffusion tensor imaging of brain development. *Semin Fetal Neonatal Med, 11*(6), 489-497.

Huttenlocher, P. R., & Dabholkar, A. S. (1997). Regional differences in synaptogenesis in human cerebral cortex. *J Comp Neurol, 387*(2), 167-178. doi: 10.1002/(SICI)1096-9861(19971020)387:2<167::AID-CNE1>3.0.CO;2-Z [pii]

Innocenti, G. M., Caminiti, R., & Aboitiz, F. (2015). Comments on the paper by Horowitz et al. (2014). *Brain Struct Funct, 220*(3), 1789-1790. doi: 10.1007/s00429-014-0974-7

Johansen-Berg, Heidi, & Behrens, Timothy EJ. (2013). *Diffusion MRI: from quantitative measurement to in vivo neuroanatomy*: Academic Press.

Judas, M., Rados, M., Jovanov-Milosevic, N., Hrabac, P., Stern-Padovan, R., & Kostovic, I. (2005). Structural, immunocytochemical, and mr imaging properties of periventricular crossroads of growing cortical pathways in preterm infants. *AJNR Am J Neuroradiol, 26*(10), 2671-2684. doi: 26/10/2671 [pii]

Kapellou, O., Counsell, S. J., Kennea, N., Dyet, L., Saeed, N., Stark, J., Maalouf, E., Duggan, P., Ajayi-Obe, M., Hajnal, J., Allsop, J. M., Boardman, J., Rutherford, M. A., Cowan, F., & Edwards, A. D. (2006). Abnormal cortical development after premature birth shown by altered allometric scaling of brain growth. *PLoS Med, 3*(8), e265. doi: 10.1371/journal.pmed.0030265

Kasprian, G., Brugger, P. C., Weber, M., Krssak, M., Krampl, E., Herold, C., & Prayer, D. (2008). In utero tractography of fetal white matter development. *Neuroimage, 43*(2), 213-224. doi: S1053-8119(08)00821-5 [pii] 10.1016/j.neuroimage.2008.07.026

Kersbergen, K. J., Leemans, A., Groenendaal, F., van der Aa, N. E., Viergever, M. A., de Vries, L. S., & Benders, M. J. (2014). Microstructural brain development between 30 and 40week corrected age in a longitudinal cohort of extremely preterm infants. *Neuroimage, 103*, 214-224. doi: S1053-8119(14)00781-2 [pii] 10.1016/j.neuroimage.2014.09.039

Kersbergen, K. J., Leroy, F., Isgum, I., Groenendaal, F., de Vries, L. S., Claessens, N. H. P., van Haastert, I. C., Moeskops, P., Fischer, C., Mangin, J. F., Viergever, M. A., Dubois, J., & Benders, Mjnl. (2016). Relation between clinical risk factors, early cortical changes, and neurodevelopmental outcome in preterm infants. *Neuroimage, 142*, 301-310. doi: 10.1016/j.neuroimage.2016.07.010

Keunen, K., Benders, M. J., Leemans, A., Fieret-Van Stam, P. C., Scholtens, L. H., Viergever, M. A., Kahn, R. S., Groenendaal, F., de Vries, L. S., & van den Heuvel, M. P. (2017). White matter maturation in the neonatal brain is predictive of school age cognitive capacities in children born very preterm. *Dev Med Child Neurol, 59*, 939-946. doi: 10.1111/dmcn.13487

Keunen, K., Counsell, S. J., & Benders, M. J. (2017). The emergence of functional architecture during early brain development. *Neuroimage, 160*, 2-14. doi: S1053-8119(17)30054-X [pii] 10.1016/j.neuroimage.2017.01.047

Kinney, H. C., Brody, B. A., Kloman, A. S., & Gilles, F. H. (1988). Sequence of central nervous system myelination in human infancy. II. Patterns of myelination in autopsied infants. *J Neuropathol Exp Neurol, 47*(3), 217-234.

Klyachko, V. A., & Stevens, C. F. (2003). Connectivity optimization and the positioning of cortical areas. *Proc Natl Acad Sci U S A, 100*(13), 7937-7941.

Knickmeyer, R. C., Gouttard, S., Kang, C., Evans, D., Wilber, K., Smith, J. K., Hamer, R. M., Lin, W., Gerig, G., & Gilmore, J. H. (2008). A structural MRI study of human brain development from birth to 2 years. *J Neurosci, 28*(47), 12176-12182. doi: 28/47/12176 [pii] 10.1523/JNEUROSCI.3479-08.2008

Kostovic, I., & Jovanov-Milosevic, N. (2006). The development of cerebral connections during the first 20-45 weeks' gestation. *Semin Fetal Neonatal Med, 11*(6), 415-422. doi: S1744-165X(06)00068-0 [pii] 10.1016/j.siny.2006.07.001

Kostovic, I., Jovanov-Milosevic, N., Rados, M., Sedmak, G., Benjak, V., Kostovic-Srzentic, M., Vasung, L., Culjat, M., Huppi, P., & Judas, M. (2014). Perinatal and early postnatal reorganization of the subplate and related cellular compartments in the human cerebral wall as revealed by histological and MRI approaches. *Brain Struct Funct, 219*(1), 231-253. doi: 10.1007/s00429-012-0496-0

Kostovic, I., & Judas, M. (2006). Prolonged coexistence of transient and permanent circuitry elements in the developing cerebral cortex of fetuses and preterm infants. *Dev Med Child Neurol, 48*(5), 388-393. doi: S0012162206000831 [pii] 10.1017/S0012162206000831

Kostovic, I., & Judas, M. (2015). Embryonic and fetal development of the human cerebral cortex. In Arthur W. Toga (Ed.), *Brain Mapping: An Encyclopedic Reference* (Vol. 2, pp. 423-437): Academic Press: Elsevier.

Kostovic, I., Kostovic-Srzentic, M., Benjak, V., Jovanov-Milosevic, N., & Rados, M. (2014). Developmental dynamics of radial vulnerability in the cerebral compartments in preterm infants and neonates. *Front Neurol, 5,* 139. doi: 10.3389/fneur.2014.00139

Kostovic, I., Petanjek, Z., & Judas, M. (1993). Early areal differentiation of the human cerebral cortex: entorhinal area. *Hippocampus, 3*(4), 447-458. doi: 10.1002/hipo.450030406

Kostovic, I., & Rakic, P. (1990). Developmental history of the transient subplate zone in the visual and somatosensory cortex of the macaque monkey and human brain. *J Comp Neurol, 297*(3), 441-470. doi: 10.1002/cne.902970309

Kroenke, C. D., & Bayly, P. V. (2018). How Forces Fold the Cerebral Cortex. *J Neurosci, 38*(4), 767-775. doi: 10.1523/JNEUROSCI.1105-17.2017

Krogsrud, S. K., Fjell, A. M., Tamnes, C. K., Grydeland, H., Mork, L., Due-Tonnessen, P., Bjornerud, A., Sampaio-Baptista, C., Andersson, J., Johansen-Berg, H., & Walhovd, K. B. (2015). Changes in white matter microstructure in the developing brain-A longitudinal diffusion tensor imaging study of children from 4 to 11years of age. *Neuroimage, 124*(Pt A), 473-486. doi: S1053-8119(15)00814-9 [pii] 10.1016/j.neuroimage.2015.09.017

Kucharczyk, W., Macdonald, P. M., Stanisz, G. J., & Henkelman, R. M. (1994). Relaxivity and magnetization transfer of white matter lipids at MR imaging: importance of cerebrosides and pH. *Radiology, 192*(2), 521-529.

Kulikova, S., Hertz-Pannier, L., Dehaene-Lambertz, G., Buzmakov, A., Poupon, C., & Dubois, J. (2015). Multi-parametric evaluation of the white matter maturation. *Brain Struct Funct, 220*(6), 3657-3672. doi: 10.1007/s00429-014-0881-y

Kulikova, S., Hertz-Pannier, L., Dehaene-Lambertz, G., Poupon, C., & Dubois, J. (2016). A New Strategy for Fast MRI-Based Quantification of the Myelin Water Fraction: Application to Brain Imaging in Infants. *PLoS One, 11*(10), e0163143. doi: 10.1371/journal.pone.0163143

Kunz, N., Zhang, H., Vasung, L., O'Brien, K. R., Assaf, Y., Lazeyras, F., Alexander, D. C., & Huppi, P. S. (2014). Assessing white matter microstructure of the newborn with multi-shell diffusion MRI and biophysical compartment models. *Neuroimage, 96*, 288-299. doi: S1053-8119(14)00218-3 [pii] 10.1016/j.neuroimage.2014.03.057

Kwan, K. Y., Sestan, N., & Anton, E. S. (2012). Transcriptional co-regulation of neuronal migration and laminar identity in the neocortex. *Development, 139*(9), 1535-1546. doi: 139/9/1535 [pii] 10.1242/dev.069963

LaMantia, A. S., & Rakic, P. (1990). Axon overproduction and elimination in the corpus callosum of the developing rhesus monkey. *J Neurosci, 10*(7), 2156-2175.

Le Guen, Y., Auzias, G., Leroy, F., Noulhiane, M., Dehaene-Lambertz, G., Duchesnay, E., Mangin, J. F., Coulon, O., & Frouin, V. (2018). Genetic Influence on the Sulcal Pits: On the Origin of the First Cortical Folds. *Cereb Cortex, 28*(6), 1922-1933. doi: 10.1093/cercor/bhx098

Lebenberg, J., Labit, M., Auzias, G., Mohlberg, H., Fischer, C., Rivière, D., Duchesnay, E., Kabdebon, C., Leroy, F., Labra, N., Poupon, F., Dickscheid, T., Hertz-Pannier, L., Poupon, C., Dehaene-Lambertz, G., Hüppi, P., Amunts, K., Dubois, J., & Mangin, J. F. (2018). A framework based on sulcal constraints to align preterm, infant and adult human brain images acquired in vivo and post mortem. *Brain Struct Funct, 223*(9), 4153-4168. doi: 10.1007/s00429-018-1735-9

Lebenberg, J., Mangin, J. F., Thirion, B., Poupon, C., Hertz-Pannier, L., Leroy, F., Adibpour, P., Dehaene-Lambertz, G., & Dubois, J. (2019). Mapping the asynchrony of cortical maturation in the infant brain: A MRI multi-parametric clustering approach. *Neuroimage, 185*, 641-653. doi: 10.1016/j.neuroimage.2018.07.022

Lebenberg, Jessica, Poupon, Cyril, Thirion, Bertrand, Leroy, François, Mangin, J-F, Dehaene-Lambertz, Ghislaine, & Dubois, Jessica. (2015). *Clustering the infant brain tissues based on microstructural properties and maturation assessment using multi-parametric MRI.* Paper presented at the 2015 IEEE 12th International Symposium on Biomedical Imaging (ISBI).

Lee, J., Birtles, D., Wattam-Bell, J., Atkinson, J., & Braddick, O. (2012). Latency measures of pattern-reversal VEP in adults and infants: different information from transient P1 response and steady-state phase. *Invest Ophthalmol Vis Sci, 53*(3), 1306-1314. doi: iovs.11-7631 [pii] 10.1167/iovs.11-7631

Leppert, I. R., Almli, C. R., McKinstry, R. C., Mulkern, R. V., Pierpaoli, C., Rivkin, M. J., & Pike, G. B. (2009). T(2) relaxometry of normal pediatric brain development. *J Magn Reson Imaging, 29*(2), 258-267. doi: 10.1002/jmri.21646

Leroy, F., Glasel, H., Dubois, J., Hertz-Pannier, L., Thirion, B., Mangin, J. F., & Dehaene-Lambertz, G. (2011). Early maturation of the linguistic dorsal pathway in human infants. *J Neurosci, 31*(4), 1500-1506. doi: 10.1523/JNEUROSCI.4141-10.2011

Li, G., Nie, J., Wang, L., Shi, F., Gilmore, J. H., Lin, W., & Shen, D. (2014). Measuring the dynamic longitudinal cortex development in infants by reconstruction of temporally consistent cortical surfaces. *Neuroimage, 90*, 266-279. doi: S1053-8119(13)01258-5 [pii] 10.1016/j.neuroimage.2013.12.038

Li, G., Nie, J., Wang, L., Shi, F., Lin, W., Gilmore, J. H., & Shen, D. (2013). Mapping region-specific longitudinal cortical surface expansion from birth to 2 years of age. *Cereb Cortex, 23*(11), 2724-2733. doi: 10.1093/cercor/bhs265

Llinares-Benadero, C., & Borrell, V. (2019). Deconstructing cortical folding: genetic, cellular and mechanical determinants. *Nat Rev Neurosci, 20*(3), 161-176. doi: 10.1038/s41583-018-0112-2

Lohmann, G., von Cramon, D. Y., & Colchester, A. C. (2008). Deep sulcal landmarks provide an organizing framework for human cortical folding. *Cereb Cortex, 18*(6), 1415-1420. doi: bhm174 [pii] 10.1093/cercor/bhm174

Lyall, A. E., Shi, F., Geng, X., Woolson, S., Li, G., Wang, L., Hamer, R. M., Shen, D., & Gilmore, J. H. (2015). Dynamic Development of Regional Cortical Thickness and Surface Area in Early Childhood. *Cereb Cortex, 25*(8), 2204-2212. doi: bhu027 [pii] 10.1093/cercor/bhu027

Maas, L. C., Mukherjee, P., Carballido-Gamio, J., Veeraraghavan, S., Miller, S. P., Partridge, S. C., Henry, R. G., Barkovich, A. J., & Vigneron, D. B. (2004). Early laminar organization of the human cerebrum demonstrated with diffusion tensor imaging in extremely premature infants. *Neuroimage, 22*(3), 1134-1140. doi: 10.1016/j.neuroimage.2004.02.035 S1053811904001429 [pii]

Makropoulos, A., Aljabar, P., Wright, R., Hüning, B., Merchant, N., Arichi, T., Tusor, N., Hajnal, J. V., Edwards, A. D., Counsell, S. J., & Rueckert, D. (2016). Regional growth and atlasing of the developing human brain. *Neuroimage, 125*, 456-478. doi: 10.1016/j.neuroimage.2015.10.047

Mangin, J. F., Jouvent, E., & Cachia, A. (2010). In-vivo measurement of cortical morphology: means and meanings. *Curr Opin Neurol, 23*(4), 359-367. doi: 10.1097/WCO.0b013e32833a0afc

Mangin, J. F., Le Guen, Y., Labra, N., Grigis, A., Frouin, V., Guevara, M., Fischer, C., Riviere, D., Hopkins, W. D., Regis, J., & Sun, Z. Y. (2019). "Plis de passage" Deserve a Role in Models of the Cortical Folding Process. *Brain Topogr, 32*(6), 1035-1048. doi: 10.1007/s10548-019-00734-8

Mangin, J. F., Lebenberg, J., Lefranc, S., Labra, N., Auzias, G., Labit, M., Guevara, M., Mohlberg, H., Roca, P., Guevara, P., Dubois, J., Leroy, F., Dehaene-Lambertz, G., Cachia, A., Dickscheid, T., Coulon, O., Poupon, C., Rivière, D., Amunts, K., & Sun, Z. Y. (2016). Spatial normalization of brain images and beyond. *Med Image Anal*. doi: 10.1016/j.media.2016.06.008

Manjón, J. V., & Coupé, P. (2016). volBrain: An Online MRI Brain Volumetry System. *Front Neuroinform, 10,* 30. doi: 10.3389/fninf.2016.00030

Matsumae, M., Kurita, D., Atsumi, H., Haida, M., Sato, O., & Tsugane, R. (2001). Sequential changes in MR water proton relaxation time detect the process of rat brain myelination during maturation. *Mech Ageing Dev, 122*(12), 1281-1291. doi: S0047-6374(01)00265-2 [pii]

Matsuzawa, J., Matsui, M., Konishi, T., Noguchi, K., Gur, R. C., Bilker, W., & Miyawaki, T. (2001). Age-related volumetric changes of brain gray and white matter in healthy infants and children. *Cereb Cortex, 11*(4), 335-342.

McCart, R. J., & Henry, G. H. (1994). Visual corticogeniculate projections in the cat. *Brain Res, 653*(1-2), 351-356. doi: 0006-8993(94)90412-X [pii]

McCulloch, D. L., Orbach, H., & Skarf, B. (1999). Maturation of the pattern-reversal VEP in human infants: a theoretical framework. *Vision Res, 39*(22), 3673-3680. doi: S0042-6989(99)00091-7 [pii]

McCulloch, D. L., & Skarf, B. (1991). Development of the human visual system: monocular and binocular pattern VEP latency. *Invest Ophthalmol Vis Sci, 32*(8), 2372-2381.

McGowan, J. C. (1999). The physical basis of magnetization transfer imaging. *Neurology, 53*(5 Suppl 3), S3-7.

McKinstry, R. C., Mathur, A., Miller, J. H., Ozcan, A., Snyder, A. Z., Schefft, G. L., Almli, C. R., Shiran, S. I., Conturo, T. E., & Neil, J. J. (2002). Radial organization of developing preterm human cerebral cortex revealed by non-invasive water diffusion anisotropy MRI. *Cereb Cortex, 12*(12), 1237-1243.

Melbourne, A., Eaton-Rosen, Z., Orasanu, E., Price, D., Bainbridge, A., Cardoso, M. J., Kendall, G. S., Robertson, N. J., Marlow, N., & Ourselin, S. (2016). Longitudinal development in the preterm thalamus and posterior white matter: MRI correlations between diffusion weighted imaging and T2 relaxometry. *Hum Brain Mapp, 37*(7), 2479-2492. doi: 10.1002/hbm.23188

Mezer, A., Yeatman, J. D., Stikov, N., Kay, K. N., Cho, N. J., Dougherty, R. F., Perry, M. L., Parvizi, J., Hua, L. H., Butts-Pauly, K., & Wandell, B. A. (2013). Quantifying the local tissue volume and composition in individual brains with magnetic resonance imaging. *Nat Med, 19*(12), 1667-1672. doi: nm.3390 [pii] 10.1038/nm.3390

Miller, S. L., Huppi, P. S., & Mallard, C. (2016). The consequences of fetal growth restriction on brain structure and neurodevelopmental outcome. *J Physiol, 594*(4), 807-823. doi: 10.1113/JP271402

Mitter, C., Prayer, D., Brugger, P. C., Weber, M., & Kasprian, G. (2015). In vivo tractography of fetal association fibers. *PLoS One, 10*(3), e0119536. doi: 10.1371/journal.pone.0119536 PONE-D-14-46515 [pii]

Monson, B. B., Eaton-Rosen, Z., Kapur, K., Liebenthal, E., Brownell, A., Smyser, C. D., Rogers, C. E., Inder, T. E., Warfield, S. K., & Neil, J. J. (2018). Differential Rates of Perinatal Maturation of Human Primary and Nonprimary Auditory Cortex. *eNeuro, 5*(1). doi: 10.1523/ENEURO.0380-17.2017 eN-NWR-0380-17 [pii]

Mota, B., & Herculano-Houzel, S. (2015). BRAIN STRUCTURE. Cortical folding scales universally with surface area and thickness, not number of neurons. *Science, 349*(6243), 74-77. doi: 10.1126/science.aaa9101

Mukherjee, P., Miller, J. H., Shimony, J. S., Conturo, T. E., Lee, B. C., Almli, C. R., & McKinstry, R. C. (2001). Normal brain maturation during childhood: developmental trends characterized with diffusion-tensor MR imaging. *Radiology, 221*(2), 349-358.

Neil, J. J., Shiran, S. I., McKinstry, R. C., Schefft, G. L., Snyder, A. Z., Almli, C. R., Akbudak, E., Aronovitz, J. A., Miller, J. P., Lee, B. C., & Conturo, T. E. (1998). Normal brain in human newborns: apparent diffusion coefficient and diffusion anisotropy measured by using diffusion tensor MR imaging. *Radiology, 209*(1), 57-66.

Neil, J., Miller, J., Mukherjee, P., & Huppi, P. S. (2002). Diffusion tensor imaging of normal and injured developing human brain - a technical review. *NMR Biomed, 15*(7-8), 543-552. doi: 10.1002/nbm.784

Ng, W. P., Cartel, N., Roder, J., Roach, A., & Lozano, A. (1996). Human central nervous system myelin inhibits neurite outgrowth. *Brain Res, 720*(1-2), 17-24. doi: 0006-8993(96)00062-5 [pii] 10.1016/0006-8993(96)00062-5

Nossin-Manor, R., Card, D., Morris, D., Noormohamed, S., Shroff, M. M., Whyte, H. E., Taylor, M. J., & Sled, J. G. (2013). Quantitative MRI in the very preterm brain: assessing tissue organization and myelination using magnetization transfer, diffusion tensor and T(1) imaging. *Neuroimage, 64*, 505-516. doi: S1053-8119(12)00920-2 [pii] 10.1016/j.neuroimage.2012.08.086

Nossin-Manor, R., Card, D., Raybaud, C., Taylor, M. J., & Sled, J. G. (2015). Cerebral maturation in the early preterm period-A magnetization transfer and diffusion tensor imaging study using voxel-based analysis. *Neuroimage, 112*, 30-42. doi: S1053-8119(15)00152-4 [pii] 10.1016/j.neuroimage.2015.02.051

O'Muircheartaigh, J., Dean, D. C., 3rd, Ginestet, C. E., Walker, L., Waskiewicz, N., Lehman, K., Dirks, H., Piryatinsky, I., & Deoni, S. C. (2014). White matter development and early cognition in babies and toddlers. *Hum Brain Mapp, 35*(9), 4475-4487. doi: 10.1002/hbm.22488

Ouyang, M., Dubois, J., Yu, Q., Mukherjee, P., & Huang, H. (2019). Delineation of early brain development from fetuses to infants with diffusion MRI and beyond. *Neuroimage, 185*, 836-850. doi: 10.1016/j.neuroimage.2018.04.017

Ouyang, M., Kang, H., Detre, J. A., Roberts, T. P. L., & Huang, H. (2017). Short-range connections in the developmental connectome during typical and atypical brain maturation. *Neurosci Biobehav Rev, 83*, 109-122. doi: S0149-7634(17)30376-7 [pii] 10.1016/j.neubiorev.2017.10.007

Ouyang, M., Liu, P., Jeon, T., Chalak, L., Heyne, R., Rollins, N. K., Licht, D. J., Detre, J. A., Roberts, T. P., Lu, H., & Huang, H. (2017). Heterogeneous increases of regional cerebral blood flow during preterm brain development: Preliminary assessment with pseudo-continuous arterial spin labeled perfusion MRI. *Neuroimage, 147*, 233-242. doi: S1053-8119(16)30755-8 [pii] 10.1016/j.neuroimage.2016.12.034

Panizzon, M. S., Fennema-Notestine, C., Eyler, L. T., Jernigan, T. L., Prom-Wormley, E., Neale, M., Jacobson, K., Lyons, M. J., Grant, M. D., Franz, C. E., Xian, H., Tsuang,

M., Fischl, B., Seidman, L., Dale, A., & Kremen, W. S. (2009). Distinct genetic influences on cortical surface area and cortical thickness. *Cereb Cortex, 19*(11), 2728-2735. doi: 10.1093/cercor/bhp026

Partridge, S. C., Mukherjee, P., Henry, R. G., Miller, S. P., Berman, J. I., Jin, H., Lu, Y., Glenn, O. A., Ferriero, D. M., Barkovich, A. J., & Vigneron, D. B. (2004). Diffusion tensor imaging: serial quantitation of white matter tract maturity in premature newborns. *Neuroimage, 22*(3), 1302-1314. doi: 10.1016/j.neuroimage.2004.02.038 S1053811904001454 [pii]

Perrot, M., Riviere, D., & Mangin, J. F. (2011). Cortical sulci recognition and spatial normalization. *Med Image Anal, 15*(4), 529-550. doi: S1361-8415(11)00030-2 [pii] 10.1016/j.media.2011.02.008

Petanjek, Z., Judas, M., Simic, G., Rasin, M. R., Uylings, H. B., Rakic, P., & Kostovic, I. (2011). Extraordinary neoteny of synaptic spines in the human prefrontal cortex. *Proc Natl Acad Sci U S A, 108*(32), 13281-13286. doi: 1105108108 [pii] 10.1073/pnas.1105108108

Petersen, S. E., & Posner, M. I. (2012). The Attention System of the Human Brain: 20 Years After. *Annu Rev Neurosci*. doi: 10.1146/annurev-neuro-062111-150525

Poduslo, S. E., & Jang, Y. (1984). Myelin development in infant brain. *Neurochem Res, 9*(11), 1615-1626.

Poh, J. S., Li, Y., Ratnarajah, N., Fortier, M. V., Chong, Y. S., Kwek, K., Saw, S. M., Gluckman, P. D., Meaney, M. J., & Qiu, A. (2015). Developmental synchrony of thalamocortical circuits in the neonatal brain. *Neuroimage, 116*, 168-176. doi: S1053-8119(15)00221-9 [pii] 10.1016/j.neuroimage.2015.03.039

Pontabry, J., Rousseau, F., Oubel, E., Studholme, C., Koob, M., & Dietemann, J. L. (2013). Probabilistic tractography using Q-ball imaging and particle filtering: Application to adult and in-utero fetal brain studies. *Med Image Anal, 17*(3), 297-310. doi: S1361-8415(12)00166-1 [pii] 10.1016/j.media.2012.11.004

Qiu, A., Fortier, M. V., Bai, J., Zhang, X., Chong, Y. S., Kwek, K., Saw, S. M., Godfrey, K., Gluckman, P. D., & Meaney, M. J. (2013). Morphology and microstructure of subcortical structures at birth: A large-scale Asian neonatal neuroimaging study. *Neuroimage, 65*, 315-323. doi: S1053-8119(12)00942-1 [pii] 10.1016/j.neuroimage.2012.09.032

Rajagopalan, V., Scott, J., Habas, P. A., Kim, K., Corbett-Detig, J., Rousseau, F., Barkovich, A. J., Glenn, O. A., & Studholme, C. (2011). Local tissue growth patterns underlying normal fetal human brain gyrification quantified in utero. *J Neurosci, 31*(8), 2878-2887. doi: 31/8/2878 [pii] 10.1523/JNEUROSCI.5458-10.2011

Rakic, P. (1988). Specification of cerebral cortical areas. *Science, 241*(4862), 170-176.

Rakic, P. (2000). Radial unit hypothesis of neocortical expansion. *Novartis Found Symp, 228*, 30-42; discussion 42-52.

Rana, S., Shishegar, R., Quezada, S., Johnston, L., Walker, D. W., & Tolcos, M. (2019). The Subplate: A Potential Driver of Cortical Folding? *Cereb Cortex, 29*(11), 4697-4708. doi: 10.1093/cercor/bhz003

Raznahan, A., Greenstein, D., Lee, N. R., Clasen, L. S., & Giedd, J. N. (2012). Prenatal growth in humans and postnatal brain maturation into late adolescence. *Proc Natl Acad Sci U S A, 109*(28), 11366-11371. doi: 1203350109 [pii] 10.1073/pnas.1203350109

Raznahan, A., Shaw, P., Lalonde, F., Stockman, M., Wallace, G. L., Greenstein, D., Clasen, L., Gogtay, N., & Giedd, J. N. (2011). How does your cortex grow? *J Neurosci, 31*(19), 7174-7177. doi: 31/19/7174 [pii] 10.1523/JNEUROSCI.0054-11.2011

Roberts, T. P., Khan, S. Y., Blaskey, L., Dell, J., Levy, S. E., Zarnow, D. M., & Edgar, J. C. (2009). Developmental correlation of diffusion anisotropy with auditory-evoked response. *Neuroreport, 20*(18), 1586-1591. doi: 10.1097/WNR.0b013e3283306854

Roell, M., Cachia, A., Matejko, A. , Houdé, O., Ansari, D. , & Borst, G. (Submitted). Sulcation of the intraparietal sulcus is related to childrens symbolic but not non-symbolic number skills

Ruoss, K., Lovblad, K., Schroth, G., Moessinger, A. C., & Fusch, C. (2001). Brain development (sulci and gyri) as assessed by early postnatal MR imaging in preterm and term newborn infants. *Neuropediatrics, 32*(2), 69-74. doi: 10.1055/s-2001-13871

Salami, M., Itami, C., Tsumoto, T., & Kimura, F. (2003). Change of conduction velocity by regional myelination yields constant latency irrespective of distance between thalamus and cortex. *Proc Natl Acad Sci U S A, 100*(10), 6174-6179. doi: 10.1073/pnas.0937380100 0937380100 [pii]

Salvan, P., Tournier, J. D., Batalle, D., Falconer, S., Chew, A., Kennea, N., Aljabar, P., Dehaene-Lambertz, G., Arichi, T., Edwards, A. D., & Counsell, S. J. (2017). Language ability in preterm children is associated with arcuate fasciculi microstructure at term. *Hum Brain Mapp, 38,* 3836-3847. doi: 10.1002/hbm.23632

Schneider, J., Kober, T., Bickle Graz, M., Meuli, R., Huppi, P. S., Hagmann, P., & Truttmann, A. C. (2016). Evolution of T1 Relaxation, ADC, and Fractional Anisotropy during Early Brain Maturation: A Serial Imaging Study on Preterm Infants. *AJNR Am J Neuroradiol, 37*(1), 155-162. doi: ajnr.A4510 [pii] 10.3174/ajnr.A4510

Schwartz, M. L., & Goldman-Rakic, P. S. (1991). Prenatal specification of callosal connections in rhesus monkey. *J Comp Neurol, 307*(1), 144-162. doi: 10.1002/cne.903070113

Shenkin, S. D., Starr, J. M., & Deary, I. J. (2004). Birth weight and cognitive ability in childhood: a systematic review. *Psychol Bull, 130*(6), 989-1013. doi: 10.1037/0033-2909.130.6.989

Sigaard, R. K., Kjaer, M., & Pakkenberg, B. (2016). Development of the Cell Population in the Brain White Matter of Young Children. *Cereb Cortex, 26*(1), 89-95. doi: bhu178 [pii] 10.1093/cercor/bhu178

Silbereis, John C, Pochareddy, Sirisha, Zhu, Ying, Li, Mingfeng, & Sestan, Nenad. (2016). The cellular and molecular landscapes of the developing human central nervous system. *Neuron, 89*(2), 248-268.

Skeide, M. A., Brauer, J., & Friederici, A. D. (2016). Brain Functional and Structural Predictors of Language Performance. *Cereb Cortex, 26*(5), 2127-2139. doi: bhv042 [pii] 10.1093/cercor/bhv042

Smyser, T. A., Smyser, C. D., Rogers, C. E., Gillespie, S. K., Inder, T. E., & Neil, J. J. (2016). Cortical Gray and Adjacent White Matter Demonstrate Synchronous Maturation in Very Preterm Infants. *Cereb Cortex, 26,* 3370-3378. doi: bhv164 [pii] 10.1093/cercor/bhv164

Stikov, N., Campbell, J. S., Stroh, T., Lavelee, M., Frey, S., Novek, J., Nuara, S., Ho, M. K., Bedell, B. J., Dougherty, R. F., Leppert, I. R., Boudreau, M., Narayanan, S., Duval, T., Cohen-Adad, J., Picard, P. A., Gasecka, A., Cote, D., & Pike, G. B. (2015). In vivo histology of the myelin g-ratio with magnetic resonance imaging. *Neuroimage, 118,* 397-405. doi: 10.1016/j.neuroimage.2015.05.023

Stiles, J., & Jernigan, T. L. (2010). The basics of brain development. *Neuropsychol Rev, 20*(4), 327-348. doi: 10.1007/s11065-010-9148-4

Takahashi, E., Folkerth, R. D., Galaburda, A. M., & Grant, P. E. (2012). Emerging cerebral connectivity in the human fetal brain: an MR tractography study. *Cereb Cortex, 22*(2), 455-464. doi: bhr126 [pii] 10.1093/cercor/bhr126

Tallinen, Tuomas, Chung, Jun Young, Rousseau, François, Girard, Nadine, Lefèvre, Julien, & Mahadevan, Lakshminarayanan. (2016). On the growth and form of cortical convolutions. *Nature Physics, 12*(6), 588.

Taylor, M. J., Menzies, R., MacMillan, L. J., & Whyte, H. E. (1987). VEPs in normal full-term and premature neonates: longitudinal versus cross-sectional data. *Electroencephalogr Clin Neurophysiol, 68*(1), 20-27.

Tissier, C., Linzarini, A., Allaire-Duquette, G., Mevel, K., Poirel, N., Dollfus, S., Etard, O., Orliac, F., Peyrin, C., Charron, S., Raznahan, A., Houde, O., Borst, G., & Cachia, A. (2018). Sulcal Polymorphisms of the IFC and ACC Contribute to Inhibitory Control Variability in Children and Adults. *eNeuro, 5*(1). doi: 10.1523/ENEURO.0197-17.2018

Travis, K. E., Curran, M. M., Torres, C., Leonard, M. K., Brown, T. T., Dale, A. M., Elman, J. L., & Halgren, E. (2014). Age-related Changes in Tissue Signal Properties Within Cortical Areas Important for Word Understanding in 12- to 19-Month-Old Infants. *Cereb Cortex, 24*, 1948-1955. doi: bht052 [pii] 10.1093/cercor/bht052

van den Heuvel, M. P., Kersbergen, K. J., de Reus, M. A., Keunen, K., Kahn, R. S., Groenendaal, F., de Vries, L. S., & Benders, M. J. (2015). The Neonatal Connectome During Preterm Brain Development. *Cereb Cortex, 25*(9), 3000-3013. doi: cercor/bhu095 [pii] 10.1093/cercor/bhu095

van den Heuvel, Martijn P, Scholtens, Lianne H, & Kahn, René S. (2019). Multi-scale neuroscience of psychiatric disorders. *Biological Psychiatry*.

Van der Knaap, M.S., & Valk, J. (1995a). Myelin and white matter. *Magnetic resonance of myelin, myelination and myelin disorders, Van der Knaap MS and Valk J eds, Springer-Verlag, Berlin*, 1-17.

Van der Knaap, M.S., & Valk, J. (1995b). Myelination and retarded myelination. *Magnetic resonance of myelin, myelination and myelin disorders, Van der Knaap MS and Valk J eds, Springer-Verlag, Berlin*.

Vasung, L., Huang, H., Jovanov-Milosevic, N., Pletikos, M., Mori, S., & Kostovic, I. (2010). Development of axonal pathways in the human fetal fronto-limbic brain: histochemical characterization and diffusion tensor imaging. *J Anat, 217*(4), 400-417. doi: JOA1260 [pii] 10.1111/j.1469-7580.2010.01260.x

Vasung, L., Lepage, C., Rados, M., Pletikos, M., Goldman, J. S., Richiardi, J., Raguz, M., Fischi-Gomez, E., Karama, S., Huppi, P. S., Evans, A. C., & Kostovic, I. (2016). Quantitative and Qualitative Analysis of Transient Fetal Compartments during Prenatal Human Brain Development. *Front Neuroanat, 10,* 11. doi: 10.3389/fnana.2016.00011

Vasung, L., Raguz, M., Kostovic, I., & Takahashi, E. (2017). Spatiotemporal Relationship of Brain Pathways during Human Fetal Development Using High-Angular Resolution Diffusion MR Imaging and Histology. *Front Neurosci, 11,* 348. doi: 10.3389/fnins.2017.00348

Von Bonin, G. (1950). *Essay on the cerebral cortex*: Springfield, IL.

Wagstyl, K., Lepage, C., Bludau, S., Zilles, K., Fletcher, P. C., Amunts, K., & Evans, A. C. (2018). Mapping Cortical Laminar Structure in the 3D BigBrain. *Cereb Cortex, 28*(7), 2551-2562. doi: 10.1093/cercor/bhy074

Walhovd, K. B., Fjell, A. M., Brown, T. T., Kuperman, J. M., Chung, Y., Hagler, D. J., Jr., Roddey, J. C., Erhart, M., McCabe, C., Akshoomoff, N., Amaral, D. G., Bloss, C. S., Libiger, O., Schork, N. J., Darst, B. F., Casey, B. J., Chang, L., Ernst, T. M., Frazier, J., Gruen, J. R., Kaufmann, W. E., Murray, S. S., van Zijl, P., Mostofsky, S., & Dale, A. M. (2012). Long-term influence of normal variation in neonatal characteristics on human brain development. *Proc Natl Acad Sci U S A, 109*(49), 20089-20094. doi: 1208180109 [pii]

10.1073/pnas.1208180109

Wee, C. Y., Tuan, T. A., Broekman, B. F., Ong, M. Y., Chong, Y. S., Kwek, K., Shek, L. P., Saw, S. M., Gluckman, P. D., Fortier, M. V., Meaney, M. J., & Qiu, A. (2017). Neonatal neural networks predict children behavioral profiles later in life. *Hum Brain Mapp, 38,* 1362-1373. doi: 10.1002/hbm.23459

Welker, W. (1988). Why does cerebral cortex fissure and fold? *Cereb Cortex, 8B,* 3-135.

Xu, G., Takahashi, E., Folkerth, R. D., Haynes, R. L., Volpe, J. J., Grant, P. E., & Kinney, H. C. (2012). Radial Coherence of Diffusion Tractography in the Cerebral White

Matter of the Human Fetus: Neuroanatomic Insights. *Cereb Cortex*. doi: bhs330 [pii] 10.1093/cercor/bhs330

Xydis, V., Astrakas, L., Zikou, A., Pantou, K., Andronikou, S., & Argyropoulou, M. I. (2006). Magnetization transfer ratio in the brain of preterm subjects: age-related changes during the first 2 years of life. *Eur Radiol, 16*(1), 215-220. doi: 10.1007/s00330-005-2796-8

Yakovlev, P. & Lecours, A. . (1967). *Regional Development of the Brain in early Life* Oxford: Blackwell Scientific.

Yakovlev, P. I. (1962). Morphological criteria of growth and maturation of the nervous system in man. *Res Publ Assoc Res Nerv Ment Dis, 39*, 3-46.

Yakovlev, P. I. and Lecours, A. R. (1967). *The myelogenetic cycles of regional maturation of the brain*. Blackwell, Oxford.

Yap, P. T., Fan, Y., Chen, Y., Gilmore, J. H., Lin, W., & Shen, D. (2011). Development trends of white matter connectivity in the first years of life. *PLoS One, 6*(9), e24678. doi: 10.1371/journal.pone.0024678 PONE-D-11-05390 [pii]

Yeatman, J. D., Wandell, B. A., & Mezer, A. A. (2014). Lifespan maturation and degeneration of human brain white matter. *Nat Commun, 5*, 4932. doi: ncomms5932 [pii] 10.1038/ncomms5932

Yu, Q., Ouyang, A., Chalak, L., Jeon, T., Chia, J., Mishra, V., Sivarajan, M., Jackson, G., Rollins, N., Liu, S., & Huang, H. (2017). Structural Development of Human Fetal and Preterm Brain Cortical Plate Based on Population-Averaged Templates. *Cereb Cortex, 26*(11), 4381-4391. doi: bhv201 [pii] 10.1093/cercor/bhv201

Zanin, E., Ranjeva, J. P., Confort-Gouny, S., Guye, M., Denis, D., Cozzone, P. J., & Girard, N. (2011). White matter maturation of normal human fetal brain. An in vivo diffusion tensor tractography study. *Brain Behav, 1*(2), 95-108. doi: 10.1002/brb3.17

Zhao, T., Xu, Y., & He, Y. (2018). Graph theoretical modeling of baby brain networks. *Neuroimage*. doi: S1053-8119(18)30545-7 [pii] 10.1016/j.neuroimage.2018.06.038

Zilles, K., Palomero-Gallagher, N., & Amunts, K. (2013). Development of cortical folding during evolution and ontogeny. *Trends Neurosci, 36*(5), 275-284. doi: 10.1016/j.tins.2013.01.006